\documentclass[preprint,aps,12pt,fleqn,oneside]{book}
\usepackage{epsf,epsfig,graphics,amsmath}
\usepackage{float}
\usepackage{makeidx}
\usepackage{latexsym}
\usepackage{amssymb}
\newcommand{\BE}{\begin{equation}}
\newcommand{\EE}{\end{equation}}
\newcommand{\BA}{\begin{array}}
\newcommand{\EA}{\end{array}}
\newcommand{\BEA}{\begin{eqnarray}}
\newcommand{\EEA}{\end{eqnarray}}

\begin{document}
\frontmatter
\thispagestyle{empty}%
{\topskip 50pt%
\begin{center}
{\large \bf Finite-size Domains in a Membrane with Two-state
Active Inclusions}

\vspace{9pt}

 {Chien-Hsun Chen}

{Department of Physics and Center for Complex Systems, National
Central University, Jhongli 32054, Taiwan}

\date{\today}
\end{center}}

\thispagestyle{plain} \topmargin -0.5 in \textheight 9.0 in

\begin{center}
{\large Abstract} \end{center}
 {\small We propose a model that
leads to the formation of non-equilibrium finite-size domains in a
biological membrane. Our model considers the active conformational
change of the inclusions and the coupling between inclusion
density and membrane curvature. Two special cases with different
interactions are studied by Monte Carlo simulations. In case (i)
exited state inclusions prefer to aggregate. In case (ii) ground
state inclusions prefer to aggregate. When the inclusion density
is not coupled to the local membrane curvature, in case (i) the
typical length scale ($\sqrt{M}$) of the inclusion clusters shows
weak dependence on the excitation rate ($K_{on}$) of the
inclusions for a wide range of $K_{on}$ but increases fast when
$K_{on}$ becomes sufficiently large; in case (ii) $\sqrt{M}\sim
{K_{on}}^{-\frac{1}{3}}$ for a wide range of $K_{on}$. When the
inclusion density is coupled to the local membrane curvature, the
curvature coupling provides the upper limit of the inclusion
clusters. In case (i) (case (ii)), the formation of the inclusions
is suppressed when $K_{off}$ ($K_{on}$) is sufficiently large such
that the ground state (excited state) inclusions do not have
sufficient time to aggregate. We also find that the mobility of an
inclusion in the membrane depends on inclusion-curvature coupling.
Our study suggests possible mechanisms that produce finite-size
domains in biological membranes. }

\chapter*{\centering Acknowledgement} This work is supported by the
National Science Council of the Republic of China under grant no.
NSC 92-2112-M-008-019, NSC 93-2112-M-008-030, and NSC
94-2112-M-008-020.  This thesis is submitted to the Department of
Physics, National Central University by Chien-Hsun Chen in partial
fulfillment of the requirements for the Master degree in Physics.
The author would like to thank Prof. Hsuan-Yi Chen for supervising
this work, and Mr. Jia-Yuan Wu and Yong-Hsiang Chen for
stimulating discussions.

\tableofcontents%



\mainmatter
\chapter{Introduction \label{chapter 1}}
\pagenumbering{arabic}%
\setcounter{page}{1}%

\begin{figure}[tbp]
\begin{center}
\rotatebox{0}{ \epsfxsize= 5.5 in \epsfbox{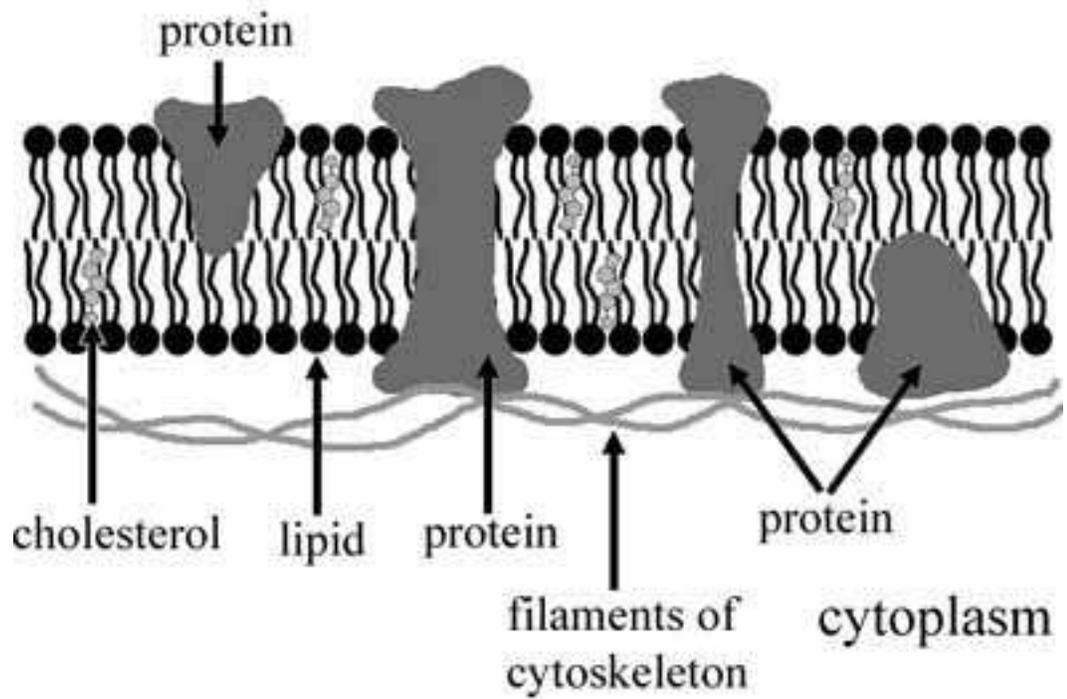} }
\caption{The configuration of membrane is shown in this figure.}
\vskip 40 pt \label{fig:introduction}
\end{center}
\end{figure}

As shown in Figure~\ref{fig:introduction}, a biological membrane
is a self-assembled bilayer composed of various types of
macromolecules such as lipids, proteins, carbohydrates,
cholesterol, and other materials.~\cite{ref:Lodish_book} A
membrane is thought to be similar to a two-dimensional fluid,
therefore the aforementioned molecules can diffuse on it. Many
experimental and theoretical studies have shown that the
distribution of these molecules in a membrane is
nonuniform.~\cite{ref:Vereb_03} Instead, some molecules aggregate
to form domains with typical length scale ranging from tens of
nanometers to about one micrometer.~\cite{ref:Veatch_04} An
important subject of current biomembrane research is to find the
mechanisms that is responsible for the formation of these
finite-size domains. From equilibrium statistical mechanism, we
know that if there are only short-range interactions between the
molecules, equilibrium macrophase separation can be induced in a
system, but equilibrium microphase separation can not be induced
if there are no long-range interactions in a system. However, up
to now there has been no clear evidence for such long-range
interactions in the membrane. Several possible mechanisms for the
formation of finite-size domains in biological membranes have been
proposed, including (i) the kinetics of spontaneous membrane
domain assembly depends on the effect of membrane recycling
ubiquitous in the living cells~\cite{ref:Matthew_05,
ref:Forest_05}, (ii) cholesterol can induce separate domains of
two mixed fatty acids ~\cite{ref:Simons_97, ref:Sparr_99}.
Although the true mechanisms of raft formation has not been
verified yet, it is possible that different mechanisms are
involved in different situations.


Since the activities of proteins in the membranes are important in
many physiological processes including ion transport, signal
transduction, et al., in this thesis we propose that the
activities of proteins can also lead to the formation of
finite-size domains in the membranes. To concentrate on the basic
mechanism behind activity-induced finite-size domains, in our
model the system contains only one type of lipids an one type of
active proteins. These proteins are called inclusions in this
thesis. The inclusions can change their internal states, and
different internal states of inclusions have different
conformations and different interaction with other molecules.
Therefore this work is a natural extension of
Ref~\cite{ref:Hsuan_04}, in which the fluctuation spectrum and the
stability of the homogeneous sate in such systems were studied.
Monte Carlo simulation is applied to illustrate our proposal. From
the result of simulation, we find that: (1) The distribution of
inclusions in a membrane can be controlled by the rate of
conformational change of the inclusions. (2) The coupling between
inclusion density and membrane curvature provides the upper limit
of the typical size of the inclusion clusters. (3) The mobility of
inclusions in the membrane depends on inclusion-curvature
coupling.

This thesis is organized as follows. In chapter 2 we introduces
the model and the Hamiltonian of our system. Chapter 3 discusses
the simulation algorithm. Chapter 4 analyzes the result of
simulation. Chapter 5 summarizes this thesis.

\setcounter{equation}{0}
\chapter{The model \label{chapter 2}}
\section{Introduction}

\begin{figure}[tbp]
\begin{center}
\rotatebox{0}{ \epsfxsize= 5.5 in \epsfbox{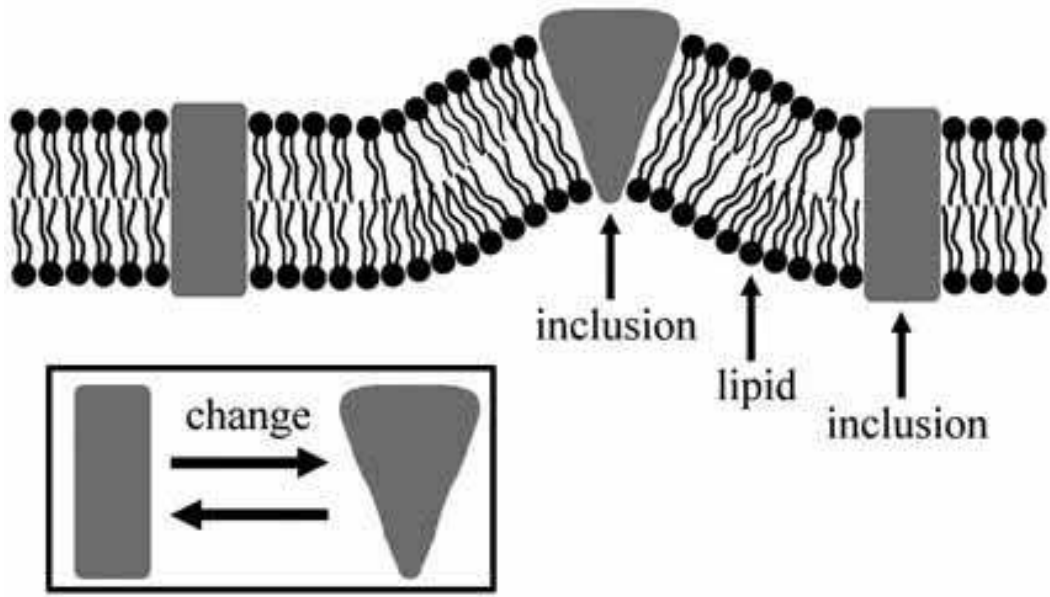} }
\caption{Schematics of two conformations of an inclusion and its
surrounding lipids. Different states of an inclusion have different
conformation, thus they induce different local membrane curvature.}
\vskip 40 pt \label{fig:changestates_01}
\end{center}
\end{figure}

Since we propose that the formation of heterogeneous structure in
biological membranes can be controlled by the activities (i.e.,
conformational changes) of the membrane inclusions, to illustrate
our proposal we consider a model in which the membrane is composed
of only one type of lipid molecules and one type of active
inclusions with two conformational states. These inclusions change
their conformational states by either external energy drive (ATP,
light, etc.), or by ligands. Therefore the membrane is called an
active membrane. Schematics of the conformations of an inclusion
and its surrounding lipids is shown in
Figure~\ref{fig:changestates_01}. Different conformations of an
inclusion prefer different local membrane curvature, thus the
inclusions also induce local membrane curvature. We discuss the
Hamiltonian that describes our model in the following section.


\section{The Hamiltonian}
\begin{figure}[tbp]
\begin{center}
\rotatebox{0}{ \epsfxsize= 5.5 in \epsfbox{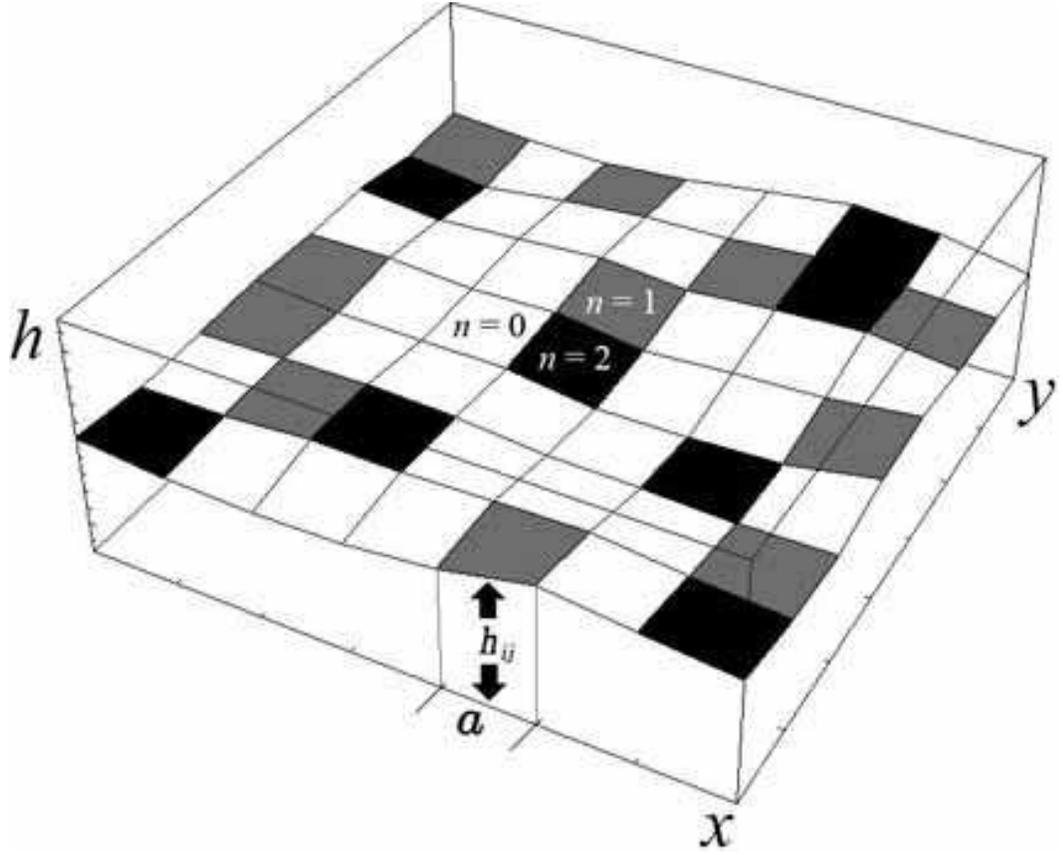} }
\caption{The configuration of the system is described by $h$ and
$n$. The reference plane is discretized into a square lattice with
lattice constant $a \sim 5nm$, and $h_{i,j}$ is the height of the
membrane measured from a flat reference plane at patch $(i,j)$. A
white patch is occupied by lipids, and $n=0$ at this patch. A gray
patch is occupied by a ground state inclusion ($n_{ij}=1$ at this
patch). A black patch is occupied by an excited state inclusion
($n_{ij}=2$ at this patch.} \vskip 40 pt \label{fig:membrane_01}
\end{center}
\end{figure}

    Our simulation is carried out on a lattice model.
    In our model, the configuration of the system is described by two
    scalar fields
    $h$ and $n$. As shown in Figure~\ref{fig:membrane_01},
    $h$ and $n$ are defined on a square lattice with lattice
    constant $a$. The size of $a$ is chosen to be the smallest
    length scale beyond which the continuum elastic description of
    membranes breaks down ($a \approx 5nm$).~\cite{ref:Weikl_02} $h_{ij}$ is the height of the
    membrane measured from a reference plane ($xy$-plane) at lattice
    patch $(i,j)$, and
    $n_{ij}$ denotes what molecule is at patch $(i,j)$ and is defined in the following way,
\begin{align}\label{eq:n_ij}
  n_{ij} = \left\{\begin{array}{ll}
  0 & \mbox{, if patch } (i,j) \mbox{ is occupied by lipids,}\\
  1 & \mbox{, if patch } (i,j) \mbox{ is occupied by an inclusion in ground state,}\\
  2 & \mbox{, if patch } (i,j) \mbox{ is occupied by an inclusion in excited
  state.}
  \end{array} \right.
\end{align}
    The lipids and
    inclusions are allowed to move in the membrane and the inclusions can change
    their conformations.

\begin{figure}[tbp]
\begin{center}
\rotatebox{0}{ \epsfxsize= 5.5 in \epsfbox{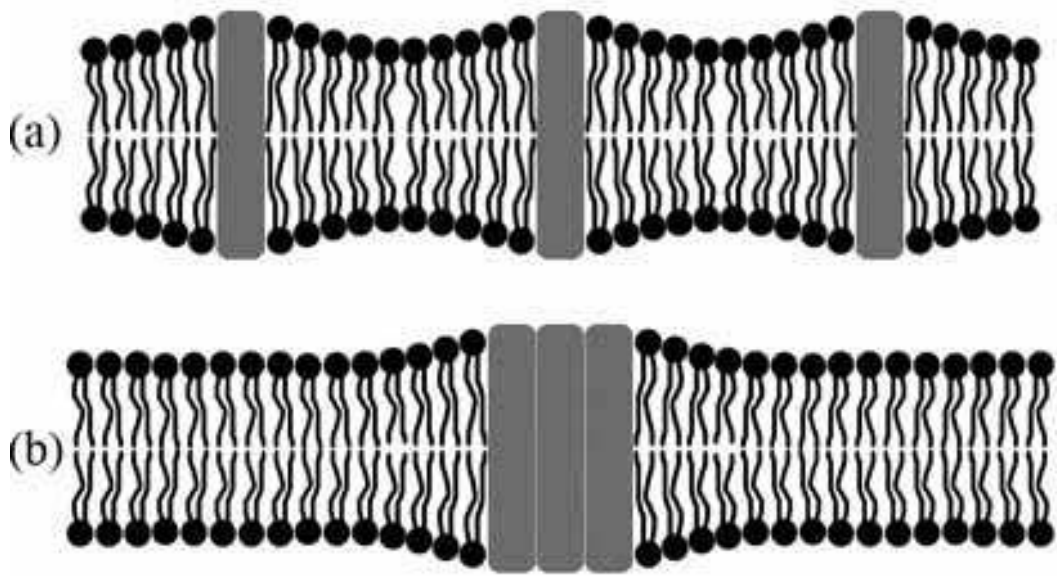} }
\caption{Inclusion-inclusion attraction can be induced by
    hydrophobic mismatch. When the hydrophobic
    length of the inclusions is different from that of a lipid
    bilayer, the lipid bilayer is stretched or compressed from its relaxed thickness.
    Then free energy of the system is increased due to this inclusion-induced thickness variation.
    Inclusions could aggregate together to lower the free energy of the system. Free energy of (b) is lower than (a).}
    \vskip 40 pt \label{fig:aggregate_01}
\end{center}
\end{figure}

    The Hamiltonian of the membrane includes: (1) the bending energy of the
    membrane, (2) the surface tension of the membrane, (3) the coupling between inclusion
    density and membrane curvature, and (4) the interaction energy
    between the inclusions and lipid molecules. Therefore the discretized form of the
    hamiltonian of the membrane that includes all the above terms can be written
    as
\begin{align}\label{eq:Hami}
  H_{lattice}=&\sum_{(i,j)}\frac{a^2}{2}\left[\kappa{\left(\nabla^2_{\bot}h\right)^2}_{ij}+\gamma\left(\left|{\nabla}_{\bot}h\right|^2\right)_{ij}\right] \notag \\
  &- \sum_{(i,j)}\sum_{p=0}^{2}a^2\kappa C_{p}\phi_{p}(i,j)\left(\nabla^2_{\bot}h\right)_{ij} \notag \\
  &+
  \frac{1}{2}\sum_{\langle(i,j)(k,l)\rangle}\left(\sum_{m,n}J_{mn}\phi_{m}(i,j)\phi_{n}(k,l)\right),
\end{align}
    where $\kappa$ is the membrane bending rigidity constant,
    and $\gamma$ is the
    surface tension constant of the membrane,
    $C_{p}$ is curvature-inclusion
    coupling of type-$p$ component, and the last term on the right hand side of
    Eq.~(\ref{eq:Hami}) is the short range interaction between
    the patches, thus $J_{mn}$ is the interaction energy between one type-$m$ patch
    and a nearest neighboring type-$n$ patch.
    The short-range interactions between the molecules in the membrane could be induced
    by hydrophobic length mismatch or other short range
    interactions.~\cite{ref:Sabra_98}
    As shown in Figure~\ref{fig:aggregate_01},
    when the hydrophobic
    length of the inclusions is different from that of a lipid
    bilayer, the lipid bilayer is stretched or compressed from its relaxed thickness.
    Then free energy of the system is increased due to this inclusion-induced thickness variation.
    Therefore the inclusions aggregate together to lower the free energy of the system.
    Besides the hydrophobic length mismatch effect, other
    short-range interactions between the inclusions also
    exist. $\sum_{(i,j)}$ means sum over all patches, and $\sum_{\langle(i,j)(k,l)\rangle}$ means sum over all
    nearest neighboring pairs. $\phi_{m}$ is defined as follows:
\begin{align}\label{eq:phi_m}
  \left\{\begin{array}{lll}
  \mbox{if } n_{ij}=0, \mbox{ then} & \phi_{0}(i,j)=1, & \phi_{1}(i,j)=\phi_{2}(i,j)=0,\\
  \mbox{if } n_{ij}=1, \mbox{ then} & \phi_{1}(i,j)=1, & \phi_{0}(i,j)=\phi_{2}(i,j)=0,\\
  \mbox{if } n_{ij}=2, \mbox{ then} & \phi_{2}(i,j)=1, &
  \phi_{0}(i,j)=\phi_{1}(i,j)=0.
  \end{array} \right.
\end{align}
   The discretized bending energy and surface energy are
\begin{equation}\label{eq:bending energy}
  \sum_{(i,j)}\frac{a^2\kappa}{2}{\left({\nabla}^2_{\bot}{h}\right)^2}_{ij}=
  \sum_{(i,j)}\frac{a^2\kappa}{2}\left(\frac{{h}_{i+1,j}+{h}_{i-1,j}-4{h}_{i,j}+{h}_{i,j+1}+{h}_{i,j-1}}{a^{2}}\right)^2,
\end{equation}
   and
\begin{equation}\label{eq:surface energy}
  \sum_{(i,j)}\frac{a^2\gamma}{2}\left(\left|{{\nabla}}_{\bot}{h}\right|^2\right)_{ij}=
  \sum_{(i,j)}\frac{a^2\gamma}{2}\left|\left(\frac{{h}_{i+1,j}-{h}_{i-1,j}}{2a}\right)\widehat{i}+\left(\frac{{h}_{i,j+1}-{h}_{i,j-1}}{2a}\right)\widehat{j}\right|^2.
\end{equation}

    Typical values of $\kappa$, $\gamma$, $a$, $h_{0}$, $C_{p}$, and $J_{mn}$ are listed in
    Table~\ref{tab:list_01}.
\begin{table}
\begin{center}
\begin{tabular}{cccc}
\hline\hline
 Symbol & Typical Value\\
\hline
 $\kappa$ & $\sim 5\times10^{-20}N\cdot m$~\cite{ref:Strey_95}\\
 $\gamma$ & $\sim 24\times10^{-6}N/m$~\cite{ref:Needham_92}\\
 $a$ & $\sim 5nm$~\cite{ref:Weikl_02}\\
 $h_{0}$ & $\sim 1nm$\\
 $C_{p}$ & $\sim 1/nm$\\
 $J_{mn}$ & $\sim 1-10k_{B}T$\\
\hline\hline
\end{tabular}\\
\end{center}
\caption{Typical values of the parameters.} \vskip 40 pt
\label{tab:list_01}
\end{table}
For convenience the non-dimensionalized hamiltonian is introduced. The unit energy is $k_{B}T$, the
    unite length in the $xy$-plane is $a$, and the unit length in
    the
    $z$-direction is
    $h_{0}\equiv a \sqrt{\frac{k_{B}T}{\kappa}}$. Therefore, the non-dimensionalized hamiltonian is
\begin{align}\label{eq:Hami_nondim}
  \widetilde{H}_{lattice}=&\sum_{(i,j)}\frac{1}{2}\left[{\left(\widetilde{\nabla}^2_{\bot}\widetilde{h}\right)^2}_{ij}
  +\frac{\gamma a^2}{\kappa}\left(\left|\widetilde{{\nabla}}_{\bot}\widetilde{h}\right|^2\right)_{ij}\right] \notag \\
  &- \sum_{(i,j)}\sum_{p=0}^{2} G_{p}\phi_{p}(i,j)\left(\widetilde{\nabla}^2_{\bot}\widetilde{h}\right)_{ij} \notag \\
  &+
  \frac{1}{2}\sum_{\langle(i,j)(k,l)\rangle}\left(\sum_{m,n}\widetilde{J}_{mn}\phi_{m}(i,j)\phi_{n}(k,l)\right),
\end{align}
    where $G_{p}\equiv h_{0}C_{p}$ is the non-dimensionalized
    inclusion-curvature coupling
    , $\widetilde{h}_{ij}\equiv\frac{1}{h_{0}}h_{ij}$,
    $\widetilde{J}_{mn}\equiv \frac{1}{k_{B}T}J_{mn}$,
    $\widetilde{{\nabla}}_{\perp}\equiv\frac{1}{a}{\nabla}_{\perp}$, and
    $\widetilde{\nabla}^2_{\perp}\equiv\frac{1}{a^2}\nabla^2_{\perp}$.

\chapter{Simulation method \label{chapter 3}}

Monte Carlo simulation is applied to study the model we introduced
in the previous chapter. In our simulation, one Monte Carlo step has
three parts: (1) The in-plane motion of materials in the membrane is
simulated by Kawasaki exchange dynamics.~\cite{ref:Binder_book} (2)
The dynamics of membrane height is simulated by Metropolis
algorithm. (3) The excitation/relaxation of the inclusions are
introduced in the simulation by assigning each inclusion certain
probability to change its state in each Monte Carlo step. When the
system has reached the steady state, Hoshen-Kopelman algorithm is
applied to analyze the size distribution of inclusion clusters in
the system.

\section{Metropolis algorithm}

Metropolis algorithm provides a set of rules to update the
microscopic state of a system such that in the long time limit the
probability that the system is in a microscopic state approaches the
equilibrium distribution. For example, consider a system described
by a single variable $x$ and equilibrium distribution $\omega(x)$.
Let us assume the system to be at state $x_{n}$ at some instance, we
first choose a trial step to change state from $x_{n}$ to $x_{t}$.
Let
\begin{align}\label{eq:Metropolis_01}
     \gamma=\omega(x_{t})/\omega(x_{n}),
\end{align}
if $\gamma\geq 1$, then the trial step is accepted and the new state
of the system become $x_{t}$. On the other hand, if $\gamma <1$,
then the trial step has probability $\gamma$ to be accepted. If the
trial step is rejected , then the system remains at state $x_{n}$.
In the following we prove that in the long time limit the
probability that the system is in state $x$ is given by
$\omega(x)$.~\cite{ref:Binder_book, ref:Gibbs_book} For an ensemble
of independent systems of random initial conditions, let $P_{n}(x)$
be the probability for a system to be in state $x$ at the $n$-th
step, and the net probability for a system changing from state $x$
to state $y$ in the next step is
\begin{align}\label{eq:Metropolis_01}
  \triangle P(x\rightarrow y)&=P_n(x)P(x\rightarrow y)-P_n(y)P(y\rightarrow
  x)\nonumber\\
  &=P_n(y)P(x\rightarrow y)\left[\frac{P_n(x)}{P_n(y)}-\frac{P(y\rightarrow x)}{P(x\rightarrow
  y)}\right].
\end{align}
Where $P(x\rightarrow y)$ is the probability for a system to change
its state from $x$ to $y$ in a step. When
\begin{align}\label{eq:Metropolis_02}
  \frac{P_n(x)}{P_n(y)}=\frac{P_e(x)}{P_e(y)}\equiv\frac{P(y\rightarrow
  x)}{P(x\rightarrow y)},
\end{align}
the system is in equilibrium. From equation
~(\ref{eq:Metropolis_01}), it is clear that if
$\frac{P_n(x)}{P_n(y)}>\frac{P_e(x)}{P_e(y)}$, then $\triangle
P(x\rightarrow y)>0$. On the other hand, when
$\frac{P_n(x)}{P_n(y)}<\frac{P_e(x)}{P_e(y)}$ then $\triangle
P(x\rightarrow y)<0$. Thus,
\begin{align}\label{eq:Metropolis_02.2}
  \lim_{n\rightarrow \infty}\frac{P_{n}(x)}{P_{n}(y)}\rightarrow \frac{P_{e}(x)}{P_{e}(y)}.
\end{align}
In the Metropolis algorithm,
\begin{align}\label{eq:Metropolis_03}
  P(x\rightarrow y)=T(x\rightarrow y)A(x\rightarrow y),
\end{align}
where $T(x\rightarrow y)$ is the probability for a trial step from
state $x$ to state $y$ to be chosen, $A(x\rightarrow y)$ is the
probability for this trial step to be accepted. In the simulation
the trial steps are chosen such that $T(x\rightarrow
y)=T(y\rightarrow x)$, therefore
\begin{align}\label{eq:Metropolis_04}
\frac{N_e(x)}{N_e(y)}=\frac{A(y\rightarrow x)}{A(x\rightarrow y)}.
\end{align}
In Metropolis algorithm, if $\omega(x)>\omega(y)$, then
$A(y\rightarrow x)=1$ and $A(x\rightarrow
y)=\frac{\omega(y)}{\omega(x)}$. If $\omega(x)<\omega(y)$, then
$A(y\rightarrow x)=\frac{\omega(y)}{\omega(x)}$ and $A(x\rightarrow
y)=1$. This leads to
\begin{align}\label{eq:Metropolis_05}
\frac{N_e(x)}{N_e(y)}=\frac{\omega(x)}{\omega(y)}.
\end{align}
That is, in the long time limit the probability distribution of the
system reaches $\omega(x)$.

\section{Monte Carlo step}
In our simulation, each Monte Carlo step has 3 parts. The first part
is Kawasaki exchange dynamics for the materials in the membrane. We
choose one patch randomly, and exchange the material on it with a
randomly chosen nearest neighbor patch with Metropolis rule. From
kawasaki exchange dynamics, time scale of one Monte Carlo step can
be estimated in the following way. The dimension of diffusion
constant is
\begin{align}\label{eq:diffusion_02}
     [D]=\frac{L^2}{T}.
\end{align}
For typical macromolecules diffusing in a membrane, $D\thicksim 1
\mu m^2/s$~\cite{ref:Weikl_04}. In the absence of any interactions,
these materials in the membrane move a distance $a\sim 5nm$ per
Monte Carlo step, therefore each Monte Carlo step corresponds to a
time interval $\Delta t\sim a^2/4D\sim 10^{-5} s$. In the second
part, we choose a new patch at random, and update the membrane
height at this patch by $\bigtriangleup h\times ran(-1,1)$ with
Metropolis algorithm, where $\bigtriangleup h$ is chosen to be
$\sim0.4nm$, and $ran(-1,1)$ is a random number between -1 and 1. In
the third part, we choose an inclusion randomly, and allow the
inclusion to change its conformation in the following way: if the
inclusion is in the ground (excited) state, then it has probability
$K_{on}\cdot \bigtriangleup t$  ($K_{off}\cdot \bigtriangleup t$) to
change its conformation to excited (ground) state, where $K_{on}$
($K_{off}$) is the excitation (relaxation) rate. The first and
second part are repeated for $N_{lat}$ (total number of patches in
the system) times per Monte Carlo step. The third part is chosen
such that on average each inclusion has performed one trial
conformational change in each Monte Carlo step. Moreover, Metropolis
algorithm is applied to the first and the second parts, but in the
third part the excitation rate and relaxation rate of inclusions are
constant. Therefore the third part drives the system out of
equilibrium, and $K_{on}$, $K_{off}$ are the important time scales
which affect the distribution of inclusions.

The size of the system in the simulations is $128\times128$ to
$256\times256$ patches. Total number of inclusions in this system is
\begin{equation}\label{eq:sumij}
    N_{inc}=\sum_{(i,j)}(\phi_{1}+\phi_{2}).
\end{equation}
For convenience, we also introduce the average inclusion density
$\phi_{inc}\equiv N_{inc}/N$. The simulations begin with
$\frac{1}{2}\phi_{inc}N$ ground state inclusions and
$\frac{1}{2}\phi_{inc}N$ excited state inclusions dispersed
randomly in a flat discretized membrane. Periodic boundary
condition is applied in all simulations. After the system has
reached steady state, the simulation is performed with up to
$10^{7}$ Monte Carlo steps, and data are taken for every 1000
Monte Carlo steps.

\section{Statistics of cluster size}
\begin{figure}[tbp]
\begin{center}
\rotatebox{0}{ \epsfxsize= 5.5 in \epsfbox{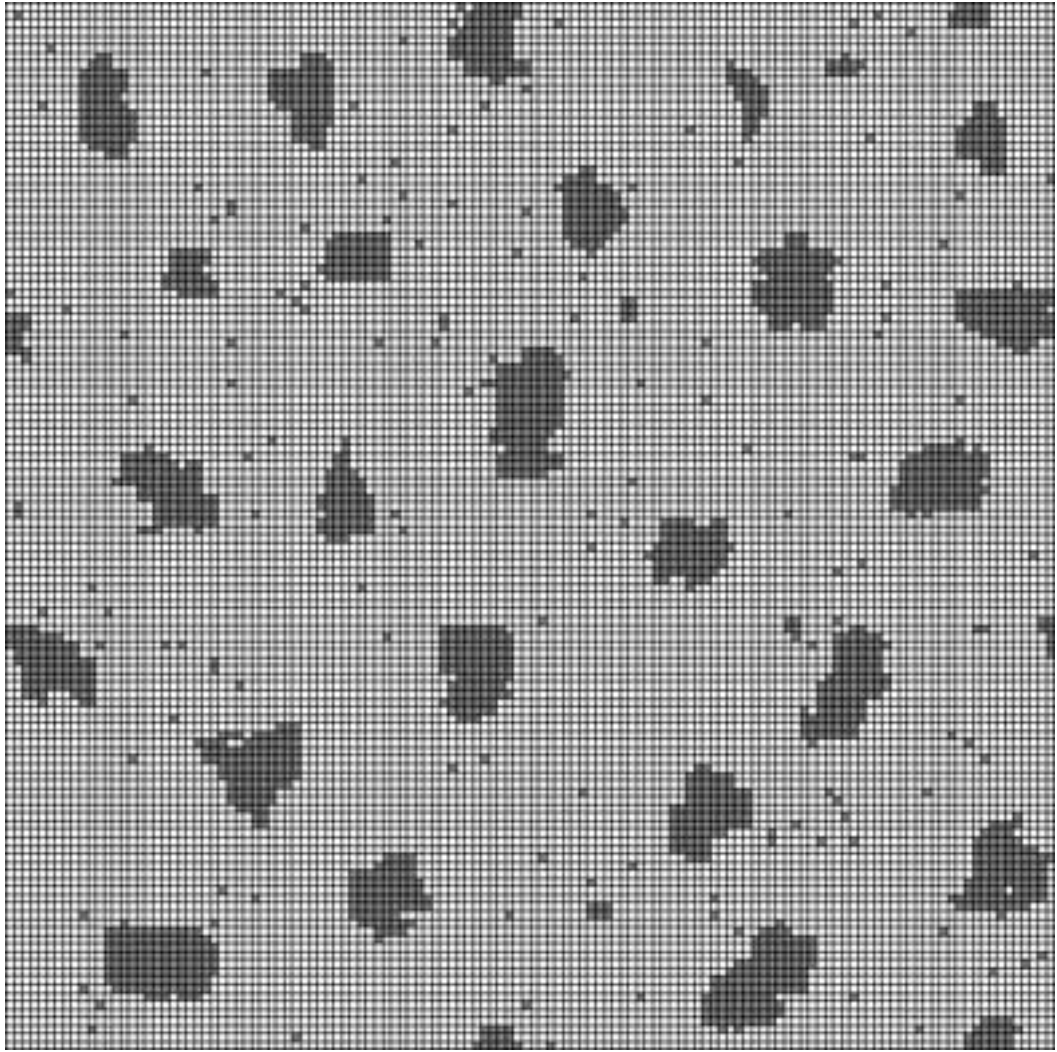} } \caption{A
$128\times128$ square lattice containing inclusions (gray patches)
and lipids (white patches).}
    \vskip 40 pt \label{fig:HK_01}
\end{center}
\end{figure}

After the system has reached steady state, Hoshen-Kopelman
algorithm, a very fast and straight method, is applied to analyze
the size distribution of inclusion
clusters.~\cite{ref:Binder_book} Figure~\ref{fig:HK_01} shows an
example of a square lattice containing inclusions (gray patches)
and lipids (white patches), and the size and number of inclusion
clusters are to be determined.

\begin{figure}[tbp]
\begin{center}
\rotatebox{0}{ \epsfxsize= 5.5 in \epsfbox{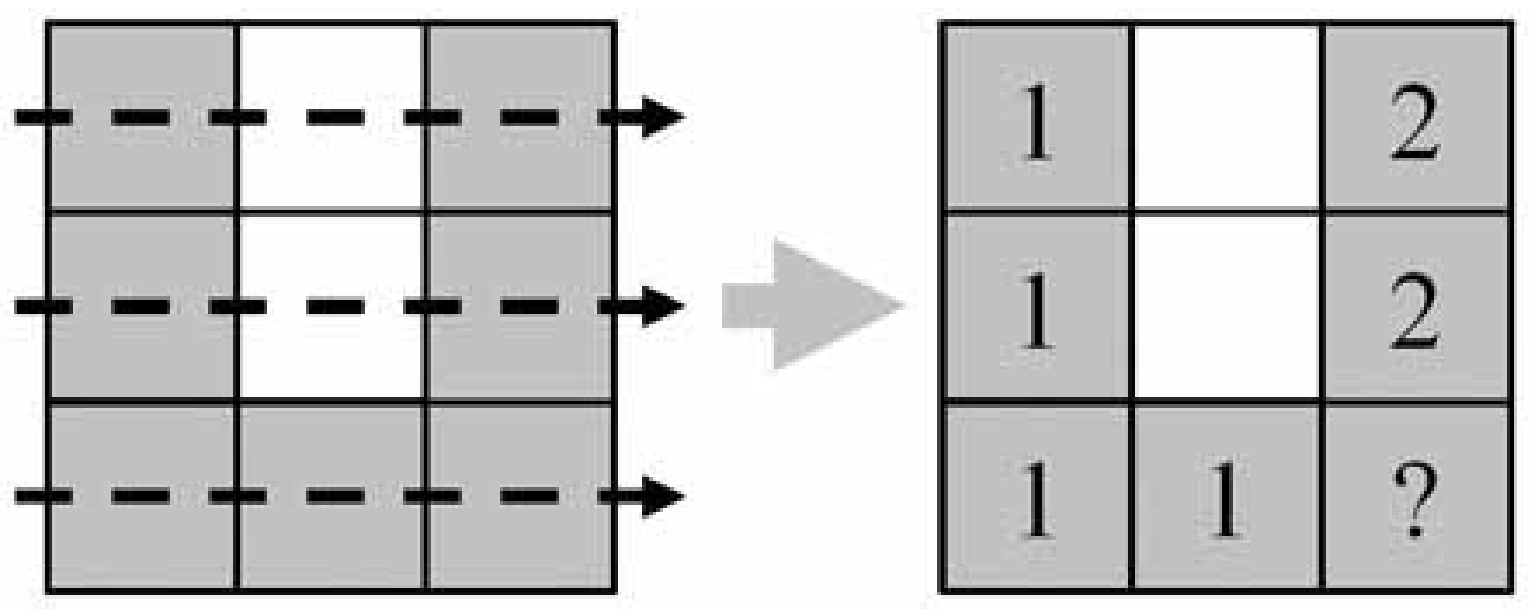} }
\caption{Labeling patches on a square lattice. This is done by going
through each row of the lattice from left to right, then from top to
bottom in turn and assigning each gray patch with a number $n$
(cluster index) in the following way. The cluster index begins from
1. If a gray patch has a nearest-neighbor gray patch that has been
visited, then the patch is assigned by the same cluster index as
this neighbor. On the other hand, if a gray patch has no
nearest-neighbor gray patch that has been visited, then we assign a
new number ($1+n_{max}$, where $n_{max}$ is the maximum $n$ that has
been assigned up to this patch) to it. In the third row, more than
one cluster indices are associated with the nearest neighbors of a
gray patch, therefore cluster 1 and cluster 2 are actually the same
cluster.}
    \vskip 40 pt \label{fig:HK_02}
\end{center}
\end{figure}

\begin{figure}[tbp]
\begin{center}
\rotatebox{0}{ \epsfxsize= 5.5 in \epsfbox{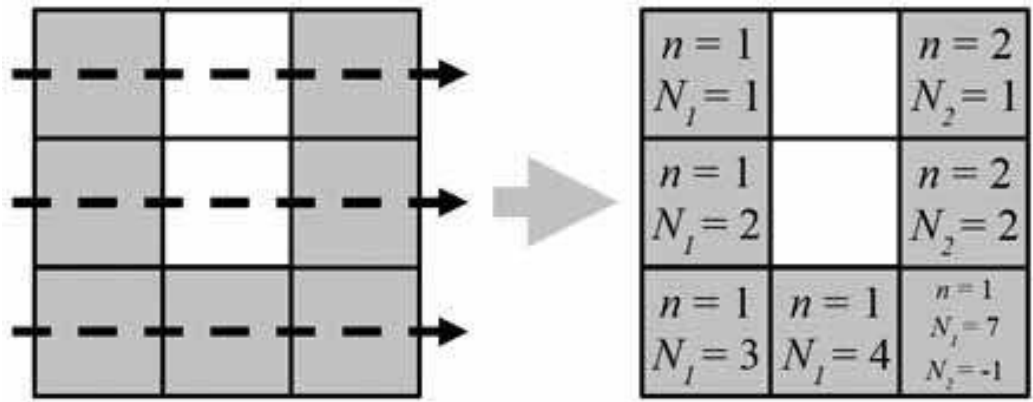} }
\caption{Hoshen-Kopelman algorithm introduces new variable $N_{n}$,
i.e., ``the labels of cluster index $n$''. If $N_{n}>0$, $N_{n}$ is
the number of patches belongs to the cluster labeled by $n$. If
$N_{n}<0$, $N_{n}$ denotes that the cluster labeled by $n$ belongs
to the cluster labeled by $-N_{n}$. In the right corner of the third
row, when there are more than one cluster index ``$n$'' associated
with the nearest neighbors of a new gray patch, we check the
$N_{n}$'s of those neighbors first. In the figure, $N_{n}$ of the
neighboring labeled patches are all positive ($N_{1}=4$, $N_{2}=2$).
Therefore for the new gray patch, n=1 (minimum cluster index of the
nearest neighbors), and $N_{1}$ is replaced by
$N_{1}+N_{2}+1=4+2+1=7$, and $N_{2}$ is replaced by -1, indicating
that cluster with label $n=2$ actually belongs to $n=1$ cluster.}
\vskip 40 pt \label{fig:HK_03}
\end{center}
\end{figure}

\begin{figure}[tbp]
\begin{center}
\rotatebox{0}{ \epsfxsize= 2.6 in \epsfbox{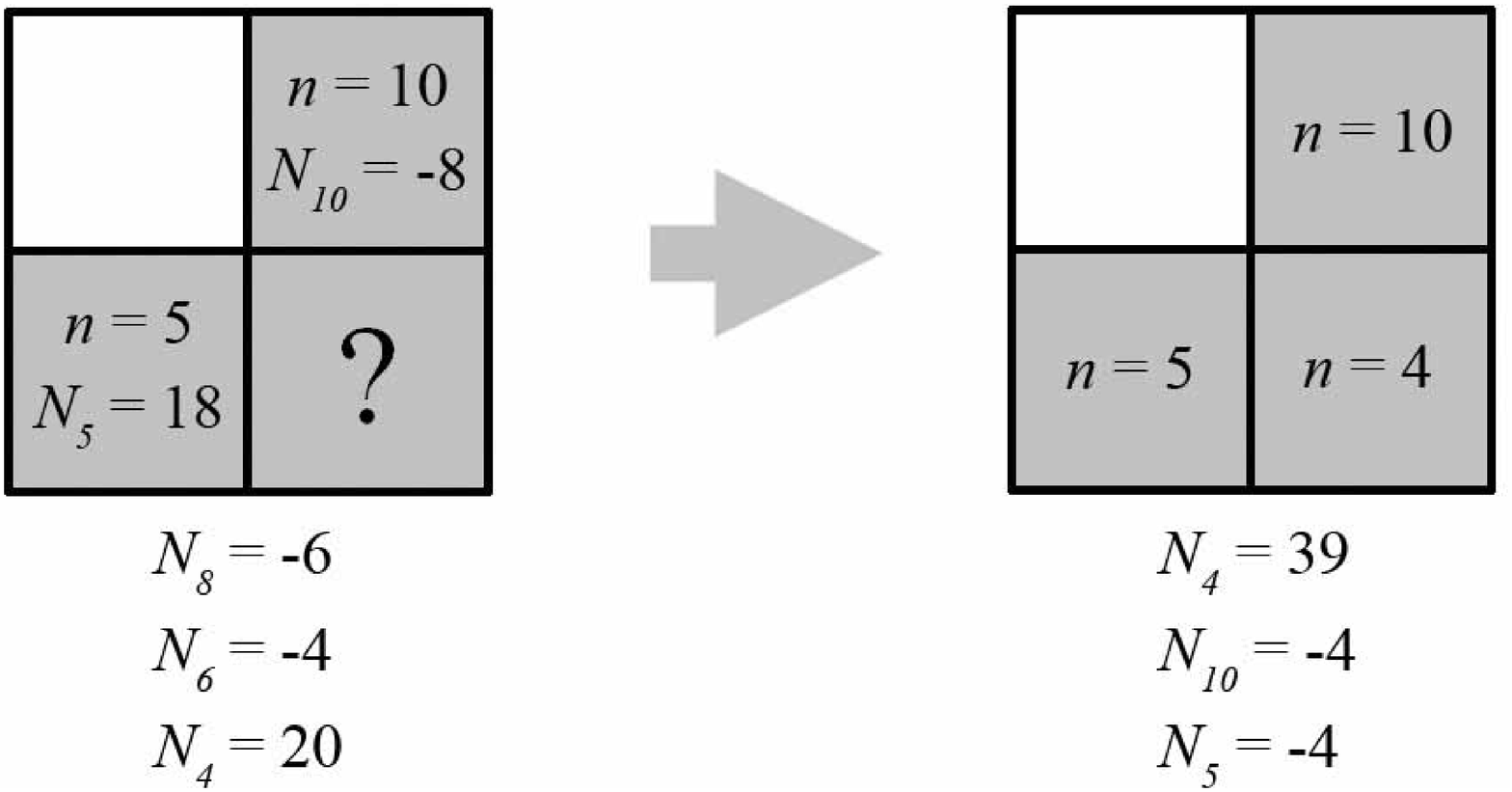} } \caption{In
this case, for the patch labeled by ``?'', $N_{n}$ of its
neighboring patches are not all positive, $N_{10}=-8$, but
$N_{8}=-6$, $N_{6}=-4$, and $N_{4}=20$. In this case, the index of
the new labeled gray patch is labeled by $n=4$, and $N_{4}$ is
replaced by $N_{4}+N_{5}+1=20+18+1=39$. $N_{10}$ and $N_{5}$ are
both replaced by -4.} \vskip 40 pt \label{fig:HK_04}
\end{center}
\end{figure}

\begin{figure}[tbp]
\begin{center}
\rotatebox{0}{ \epsfxsize= 2.6 in \epsfbox{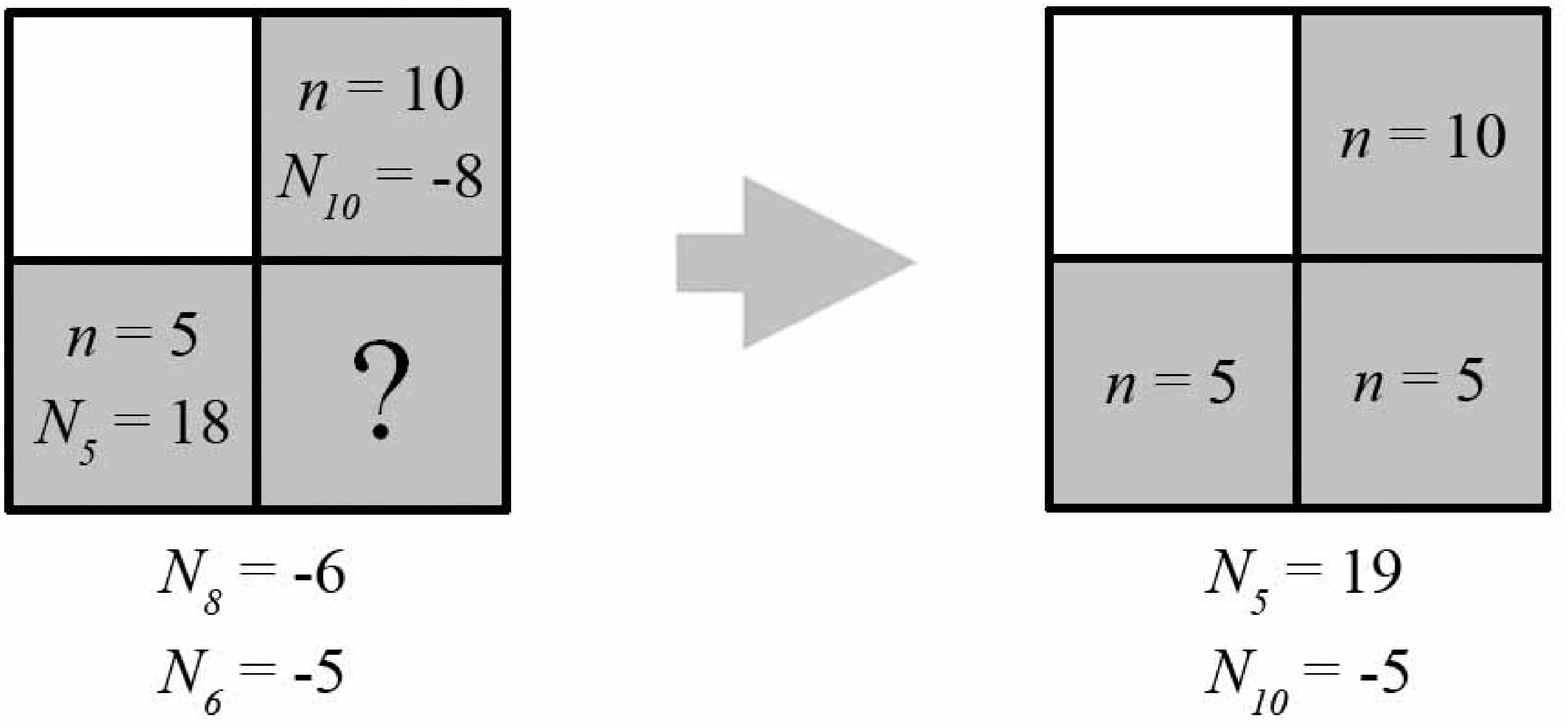} } \caption{In
this case, $N_{10}=-8$, $N_{8}=-6$, $N_{6}=-5$, and $N_{5}=18$, thus
the neighbors of the patch labeled by ``?'' actually belong to the
same cluster. The index of this new patch is $n=5$, and $N_{5}$ is
replaced by $N_{5}+1=18+1=19$, and $N_{10}$ is replaced by -5.}
\vskip 40 pt \label{fig:HK_05}
\end{center}
\end{figure}

As shown in Figure~\ref{fig:HK_02}, this is done by going through
each row of the lattice from left to right, from top to bottom in
turn and assigning each gray patch with a number $n$ which we call
``cluster index'' in the following way.
The cluster index begins from 1. When a gray patch has a
nearest-neighbor gray patch that has been visited, then the patch
is assigned by the same cluster index as this neighbor. On the
other hand, when a gray patch has no nearest-neighbor gray patch
that has been visited, then we assign a new number ($1+n_{max}$,
where $n_{max}$ is the maximum $n$ that has been assigned up to
this patch) to this patch. More than one cluster index can be
associated with the visited nearest-neighbors of a gray patch.
Figure~\ref{fig:HK_02}-\ref{fig:HK_05} illustrate some examples.
The example in Figure~\ref{fig:HK_02} is when cluster 1 and
cluster 2 are actually the same cluster. In this situation, the
Hoshen-Kopelman method corrects such ``mislabeling'' by
introducing another set of variable $N_{n}$, which is called ``the
label of the $n^{th}$ cluster''. If $N_{n}>0$, $N_{n}$ keeps track
of the number of patches belong to the $n^{th}$ cluster. If
$N_{n}<0$, $N_{n}$ denotes that the $n^{th}$ cluster actually
belongs to the cluster labeled by $-N_{n}$.

The way to determine the cluster index in Figure~\ref{fig:HK_02}
is discussed in Figure~\ref{fig:HK_03}.
In Figure~\ref{fig:HK_04}, the $n$'s and $N_{n}$'s of the nearest
neighbors of the new gray patch are different, and they belong to
different clusters. In Figure~\ref{fig:HK_05}, the $n$'s and
$N_{n}$'s of the nearest neighbors of the new gray patch are again
different, but they actually belong to the same cluster.
Figure~\ref{fig:HK_04} and~\ref{fig:HK_05} explain how
Hoshen-Kopelman method update both $n$ and $N_{n}$. After labeling
all patches of the system, we obtain the number of clusters and the
distribution of cluster size from the final values of $N_{n}$'s.

\setcounter{equation}{0}
\chapter{Simulation results and discussion \label{chapter 4}}
\section{Cluster size distribution}

\begin{figure}[tbp]
\begin{center}
\rotatebox{0}{ \epsfxsize= 4.5 in \epsfbox{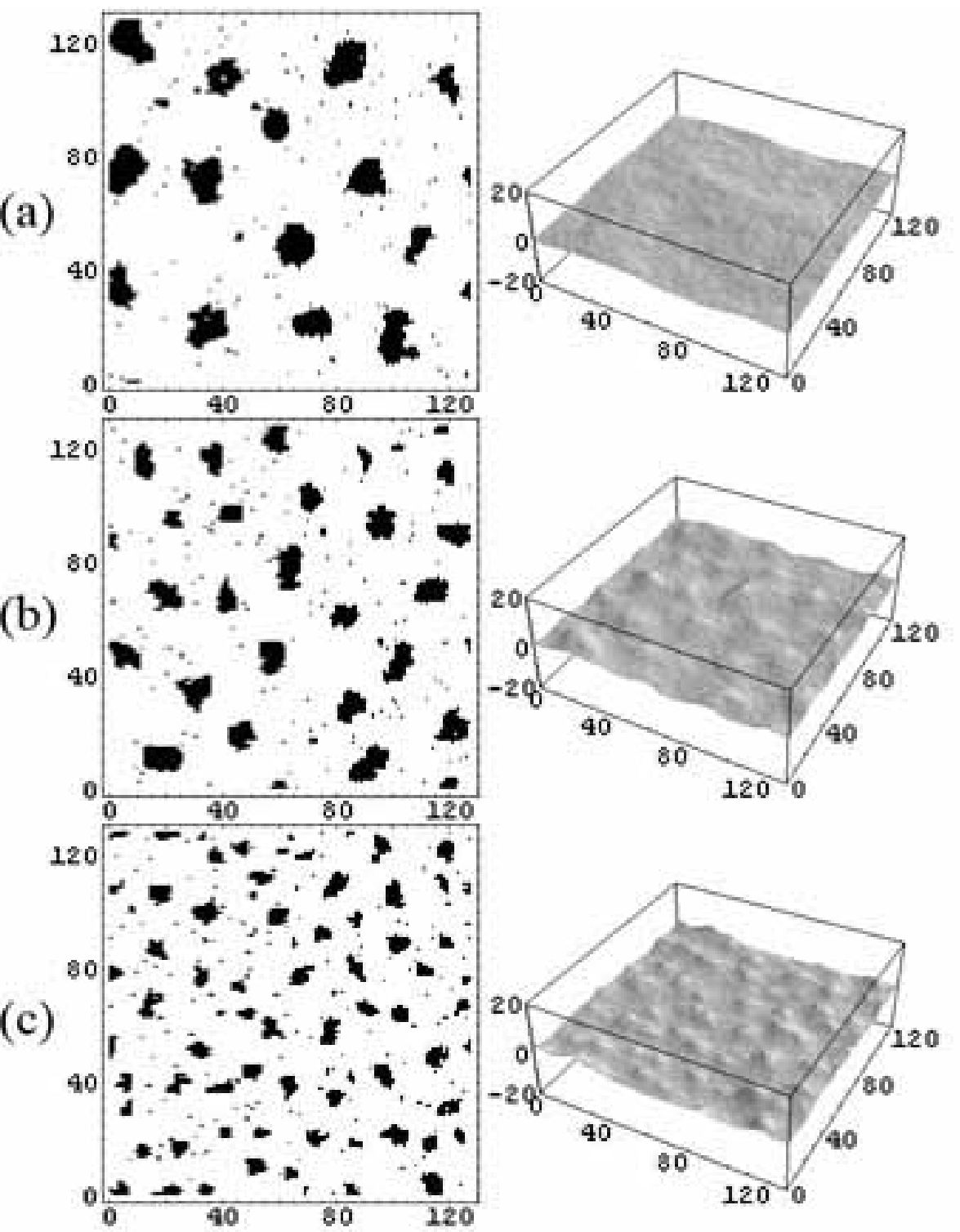} }
\caption{Typical snapshots for
     steady state inclusion distribution and membrane morphology
     for $\kappa=5\times10^{-20}N\cdot m$, $\gamma=24\times10^{-6}N/m$, $\phi_{inc}=12.5\%$, $K_{off}\Delta t=10^{-3}$,
     $K_{on}=K_{off}/32$, $\widetilde{J}_{01}=\widetilde{J}_{12}=\widetilde{J}_{0}+1.5$, all other
     $\widetilde{J}_{mn}=\widetilde{J}_{0}$, and $G_{2}=0$. (a) $G_{1}=0$. (b) $G_{1}=-1$. (c) $G_{1}=-2$. As
     $|G_{1}|$ increases, the curvature of the membrane close to the inclusion-rich domains increases and
     the typical size of inclusion clusters becomes smaller. The length scale in the $z$-direction
     is chosen to be the same as $x$ and $y$ directions (the unit length is $a$) in order to
     faithfully present the morphology of the membrane.} \vskip 40 pt \label{fig:ex_mem_cluster}
\end{center}
\end{figure}

     The simulations begin with
     $\frac{1}{2}\phi_{inc}N$ ground state inclusions and
     $\frac{1}{2}\phi_{inc}N$ excited state inclusions dispersed
     randomly in a flat discretized membrane. Periodic boundary
     condition is applied in all simulations. After the system has
     reached steady state, the simulation is performed with up to
     $10^{7}$ Monte Carlo steps, and data are taken for every 1000
     Monte Carlo steps. The size of all inclusion clusters can be found
     by the method discussed in Chapter 3. The distribution of inclusion cluster
     size is defined by
\begin{align}\label{eq:def_P}
     P(\omega)=\omega \cdot\left[\frac{n_{\omega}}{\sum_{\omega=1}^{\infty}(\omega \cdot
     n_{\omega})}\right],
\end{align}
     where $P(\omega)$ is the probability that an inclusion is found in a cluster with $\omega$ patches, and $n_{\omega}$
     is the total number of clusters with $\omega$ patches.
     Figure~\ref{fig:32_0,1,2} shows $P(\omega)$ for
     $\kappa=5\times10^{-20}N\cdot m$, $\gamma=24\times10^{-6}N/m$,
     $\phi_{inc}=12.5\%$, $K_{off}\Delta t=10^{-3}$,
     $K_{on}/K_{off}=1/32$, $\widetilde{J}_{01}=\widetilde{J}_{12}=\widetilde{J}_{0}+1.5$, all other
     $\widetilde{J}_{mn}=\widetilde{J}_{0}$, $G_{2}=0$, $G_{1}=0$, $-1$, and $-2$.
\begin{figure}[tbp]
\begin{center}
\rotatebox{0}{ \epsfxsize= 5.5 in \epsfbox{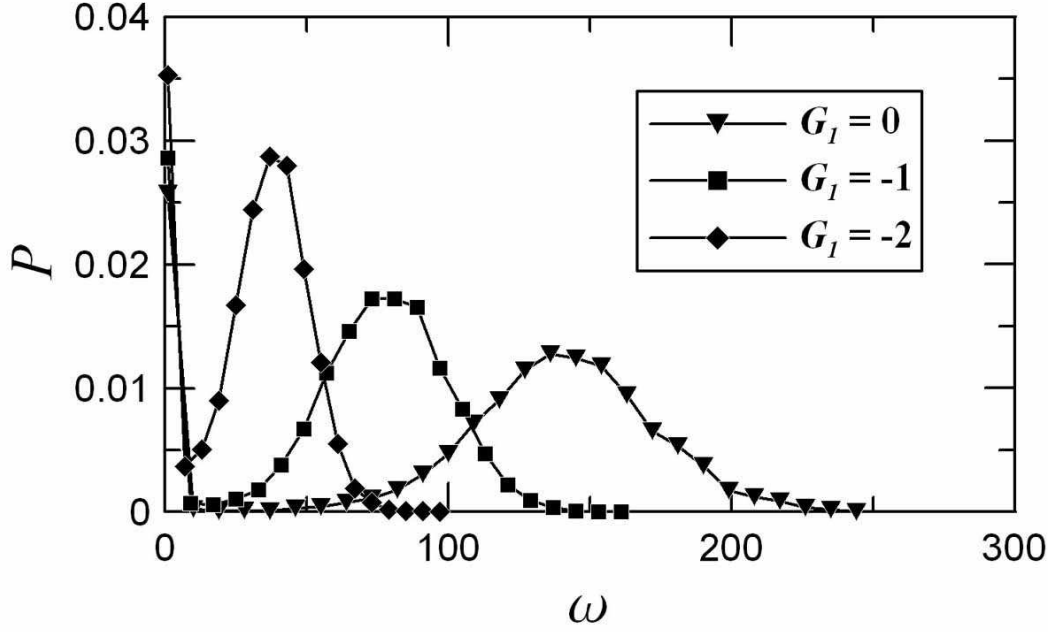} }
\caption{$P(\omega)$, the probability that an inclusion is found in
a cluster with $\omega$-patches, for
     $\kappa=5\times10^{-20}N\cdot m$,
     $\gamma=24\times10^{-6}N/m$, $\phi_{inc}=12.5\%$, $K_{off}\Delta
     t=10^{-3}$,
     $K_{on}=K_{off}/32$,
     $\widetilde{J}_{01}=\widetilde{J}_{12}=\widetilde{J}_{0}+1.5$, all other
     $\widetilde{J}_{mn}=\widetilde{J}_{0}$, $G_{2}=0$, and $G_{1}=0$ (triangle-down),
     $G_{1}=-1$ (square), $G_{1}=-2$ (diamond). $P(\omega)$ has maximum at $\omega=1$
     comes from single-inclusion clusters
     spreading around in the ``inclusion-poor'' domains. Another maximum of $P(\omega)$
     at greater $M$ comes from finite-size inclusion clusters, this
     is the typical size of the inclusion clusters and it decreases as $|G_{1}|$
     increases due to inclusion-curvature coupling.} \vskip
40 pt \label{fig:32_0,1,2}
\end{center}
\end{figure}
     The maximum at $\omega=1$ comes from single-inclusion clusters
     spreading around in the ``inclusion-poor'' domains. Another maximum of $P(\omega)$
     at greater $\omega$ comes from finite-size inclusion clusters, this
     is the typical size of the inclusion clusters.
     Figure~\ref{fig:ex_mem_cluster} shows typical snapshots for
     steady state inclusion distribution and membrane morphology
     for the same parameters as in Figure~\ref{fig:32_0,1,2}.
     The length scale in the $z$-direction
     is chosen to be the same as $x$ and $y$ directions (the unit length is $a$) in order to
     faithfully present the morphology of the membrane. One clearly sees that as
     $|G_{1}|$ increases the curvature of the membrane in the inclusion-rich domains increases and
     the typical size of inclusion clusters becomes smaller.
     Furthermore, when $G_{1}\neq0$ the location of inclusion
     clusters have strong correlation with the regions with high local membrane
     curvature. This is because the system has lower free energy when the
     inclusions locate in the regions with greater membrane
     curvature. Therefore, when $G\neq0$, the membrane prefers to form many
     mountain-like regions with inclusions aggregating on
     them.

\section{Simulation results}
To focus on the mechanisms that we propose in our model clearly,
     two special cases with different interactions between the
     inclusions and lipids are studied in the simulations.

\begin{figure}[tbp]
\begin{center}
\rotatebox{0}{ \epsfxsize= 6 in \epsfbox{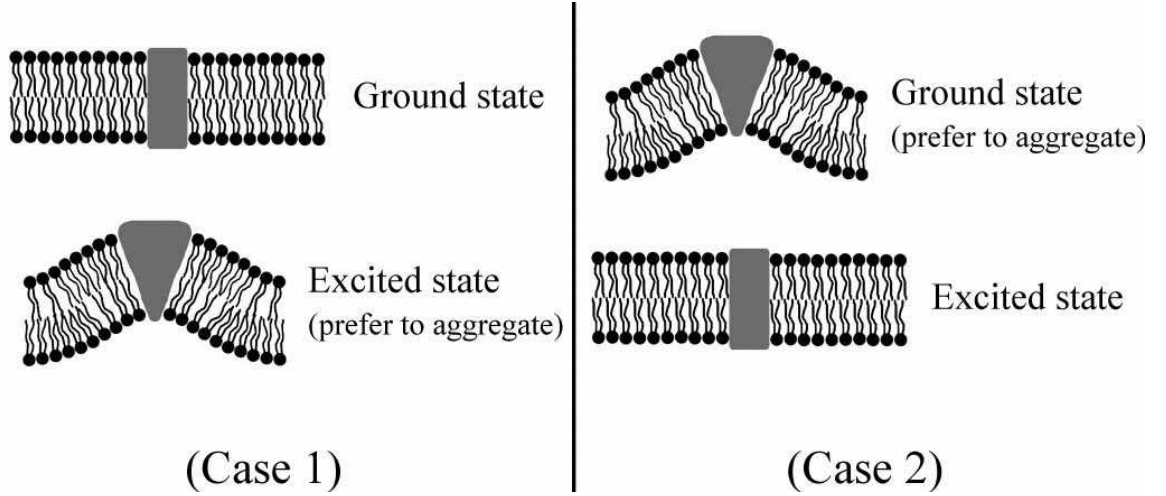} }
\caption{Schematics of inclusion conformation and
         inclusion-curvature coupling. In case 1, the excited state
         inclusions induce local membrane curvature, and the density
         of ground state inclusions is not coupled to local membrane
         curvature. In case 2, the ground state
         inclusions induce local membrane curvature, and the density
         of excited state inclusions is not coupled to local membrane
         curvature.}
\vskip 40 pt \label{fig:2case_01}
\end{center}
\end{figure}

     In case 1, $\widetilde{J}_{02}=\widetilde{J}_{12}\equiv \widetilde{J}_{0}+\widetilde{J}>0$, $\widetilde{J}_{00}=\widetilde{J}_{11}=\widetilde{J}_{22}=\widetilde{J}_{01}=\widetilde{J}_{0}$,
     $G_{1}=0$, $G_{2}\neq0$, i.e., effectively excited state
     inclusions tend to attract with each other and
     induce local membrane curvature; ground
     state inclusions do not attract or repel with other molecules, and the density
     of ground state inclusions is not coupled to local membrane
     curvature. In case 2, $\widetilde{J}_{01}=\widetilde{J}_{12}\equiv \widetilde{J}_{0}+\widetilde{J}>0$, $\widetilde{J}_{00}=\widetilde{J}_{11}=\widetilde{J}_{22}=\widetilde{J}_{02}=\widetilde{J}_{0}$,
     $G_{2}=0$, $G_{1}\neq0$, i.e., effectively ground state
     inclusions tend to attract with each other and
     induce local membrane curvature; excited
     state inclusions do not attract or repel with other molecules, and the density
     of ground state inclusions is not coupled to local membrane
     curvature. Schematics of the conformations of inclusions are
     shown in Figure~\ref{fig:2case_01}.
     We fix $K_{off}$, the relaxation rate of excited inclusions, and study
     the distribution of inclusions for different $\phi_{inc}$, $K_{on}$, $G_{1}$, and $G_{2}$.
     All simulations are carried out with $\kappa=5\times10^{-20}N\cdot m$,
     $\gamma=24\times10^{-6}N/m$, $a=5nm$, $h_{0}=1nm$, and $\Delta t \approx 10^{-5}s$.

\subsection{Short-time in-plane motions of inclusions}
\begin{figure}[tbp]
\begin{center}
\rotatebox{0}{ \epsfxsize= 5.5 in \epsfbox{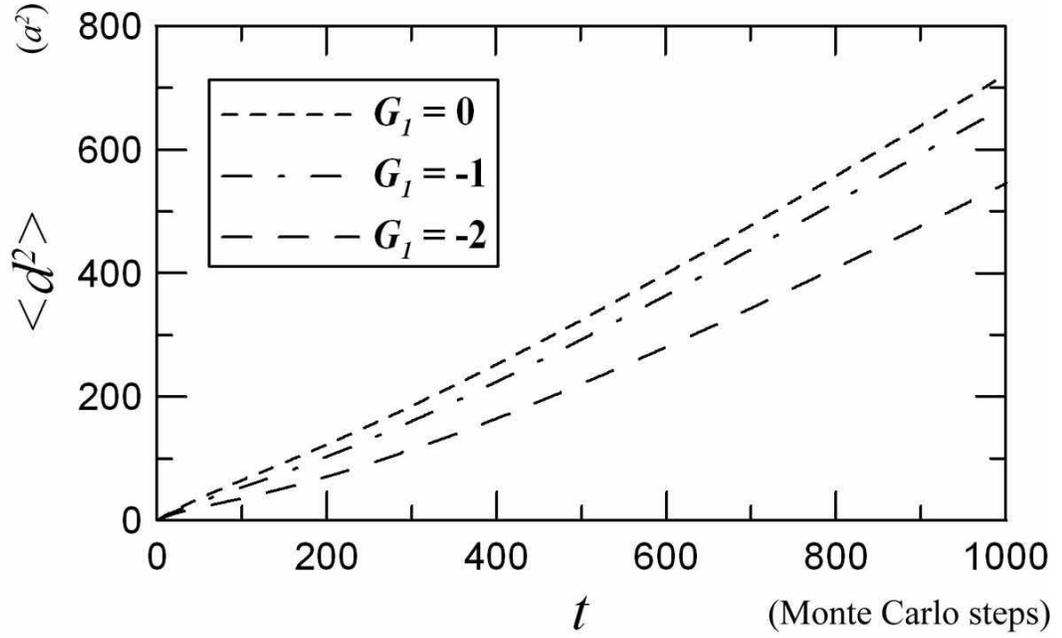} }
\caption{$\langle d^{2} \rangle$ of an inclusion within short time
interval $t$ for case 2 with $\phi_{inc}=12.5\%$,
$\widetilde{J}=1.5$, $K_{off}\Delta t=10^{-3}$,
$K_{on}=K_{off}/128$, and $G_{2}=0$. $\langle d^2\rangle\sim t$ at
sufficiently large $t$, and the inclusions with stronger
inclusion-curvature coupling move slower in the membrane.} \vskip 40
pt \label{fig:125-g-128-case}
\end{center}
\end{figure}

\begin{figure}[tbp]
\begin{center}
\rotatebox{0}{ \epsfxsize= 5.5 in \epsfbox{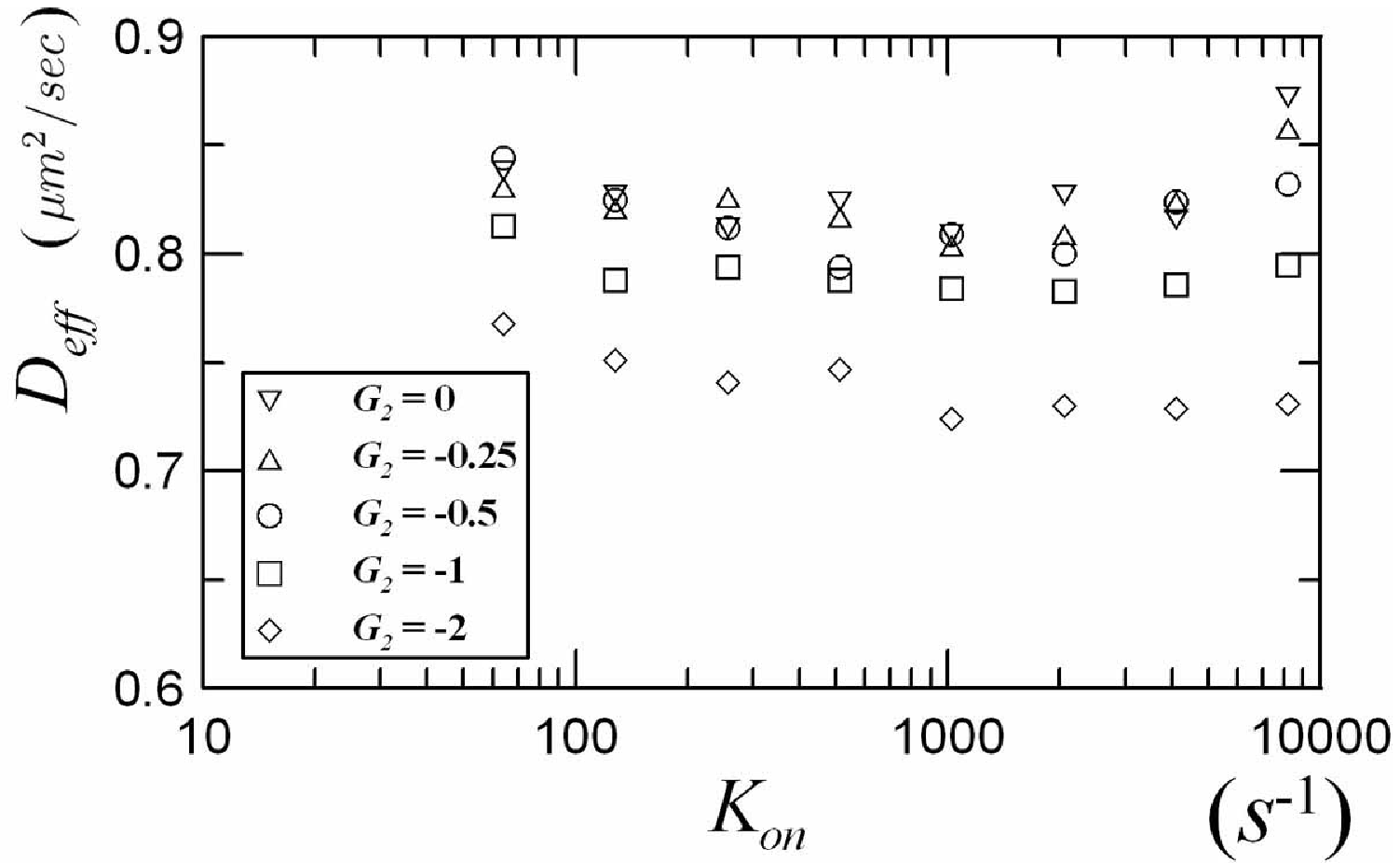} }
\caption{$D_{eff}$ in case 1 with $\phi_{inc}=12.5\%$,
$\widetilde{J}=1.5$, $K_{off}\Delta t=10^{-4}$, and $G_{2}=0$
(triangle-down), $-0.25$ (triangle-up), $-0.5$ (circle), $-1$
(square), and $-2$ (diamond). We can find that $D_{eff}$ decreases
as $|G_{2}|$ increases, but $D_{eff}$ does not depend strongly on
$K_{on}$. At large $|G_{2}|$, the tending is that $D_{eff}$ first
decreases as $K_{on}$ increases, then the dependence becomes less
significant at large $K_{on}$.} \vskip 40 pt
\label{fig:125-e-diffusion}
\end{center}
\end{figure}

\begin{figure}[tbp]
\begin{center}
\rotatebox{0}{ \epsfxsize= 5.5 in \epsfbox{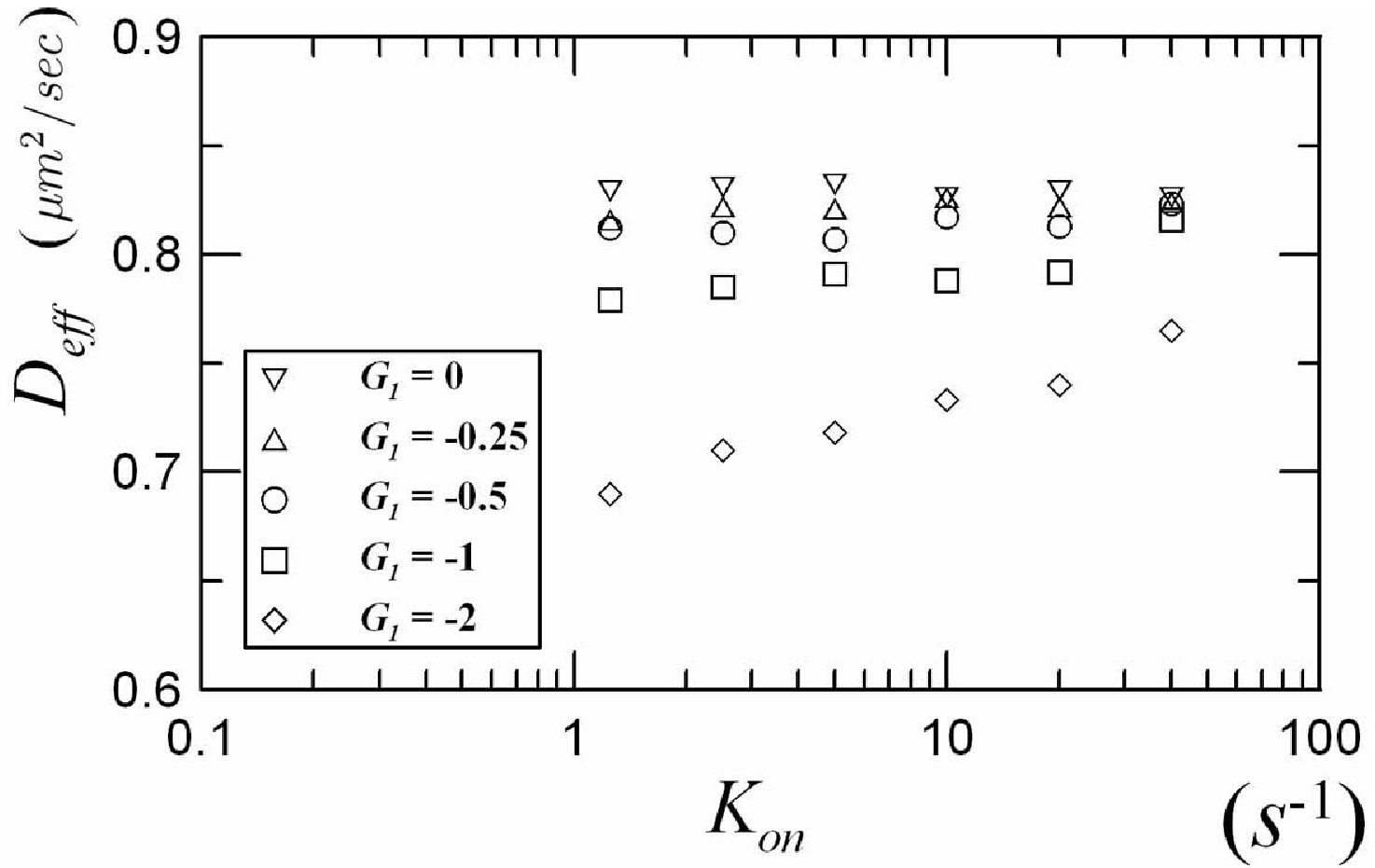} }
\caption{$D_{eff}$ in case 2 with $\phi_{inc}=12.5\%$,
$\widetilde{J}=1.5$, $K_{off}\Delta t=10^{-3}$, and $G_{1}=0$
(triangle-down), $-0.25$ (triangle-up), $-0.5$ (circle), $-1$
(square), and $-2$ (diamond). It is clear that $D_{eff}$ decreases
when $|G_{2}|$ increases. Another interesting feature is that when
$G_{1}=-2$, $D_{eff}$ shows strong dependence on $K_{on}$.} \vskip
40 pt \label{fig:125-g-diffusion}
\end{center}
\end{figure}

Although the main focus of this thesis is the size distribution of
inclusion clusters, the effects of conformational change of the
inclusions and inclusion-curvature coupling to the motion of the
inclusions in the membrane are also of great interest. Therefore
we begin our discussion with short-time in-plan motion of the
inclusions. The mean square displacement $\langle d^2(t)\rangle$
of an inclusion in short time interval $t$ for case 2 with
$\phi_{inc}=12.5\%$, $\widetilde{J}=1.5$, $K_{off}\Delta
t=10^{-3}$, and $K_{on}=K_{off}/128$ is illustrated in
Figure~\ref{fig:125-g-128-case}.
It is clear that $\langle d^2\rangle\sim t$ at sufficiently large
$t$, and the inclusions with greater inclusion-curvature coupling
move slower in the membrane. Thus we define ``effective diffusion
constant'' of an inclusion as the slope of the straight region of
the $\langle d^{2}\rangle - t$ curve.
Figure~\ref{fig:125-e-diffusion} shows the relation between
$D_{eff}$ and $K_{on}$ for case 1 with $\phi_{inc}=12.5\%$,
$\widetilde{J}=1.5$, and $K_{off}\Delta t=10^{-4}$.
It is clear that $D_{eff}$ decreases when $|G_{2}|$ increases, but
$D_{eff}$ does not depend strongly on $K_{on}$.
Figure~\ref{fig:125-g-diffusion} shows the relation between
$D_{eff}$ and $K_{on}$ in case 2 with $\phi_{inc}=12.5\%$,
$\widetilde{J}=1.5$, and $K_{off}\Delta t=10^{-3}$.
It is clear that $D_{eff}$ decreases when $|G_{1}|$ increases, too.
Another interesting feature is that when $G_{1}=-2$, $D_{eff}$ shows
strong dependence on $K_{on}$.

\begin{figure}[tbp]
\begin{center}
\rotatebox{0}{ \epsfxsize= 5.5 in \epsfbox{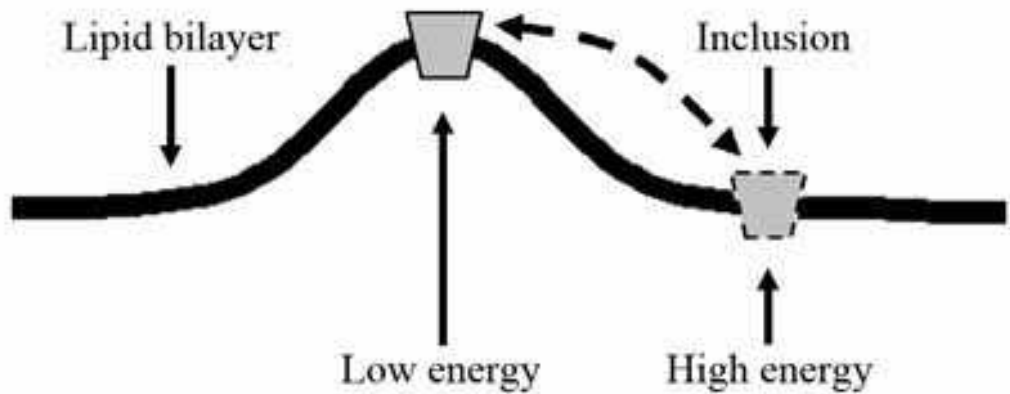} }
\caption{The strong dependence of $D_{eff}$ on $|G|$ is explained in
this figure. The free energy is lower when a curved inclusion stays
in a curved region, and the free energy is higher when a curved
inclusion stays in a flat region. Therefore it is less possible for
this inclusion to move from a curved region to a flat region. When
the inclusion-curvature coupling is stronger, this ``caging'' effect
is more significant.} \vskip 40 pt \label{fig:in_plan_motion}
\end{center}
\end{figure}

Figure~\ref{fig:in_plan_motion} explains the strong dependence of
$D_{eff}$ on $|G|$. The free energy is lower when a curved
inclusion stays in a curved region, and the free energy is higher
when a curved inclusion stays in a flat region.
Therefore it is less possible for this inclusion to move from a
curved region to a flat region. When the inclusion-curvature
coupling is stronger, this ``caging'' effect is more significant.
This explains why large inclusion-curvature coupling results in
small $D_{eff}$.

Moreover, in case 2, when $K_{on}$ increases an inclusion spends
less time in the ground state. Thus $D_{eff}$ strongly depends on
the $K_{on}$ when $|G_{1}|$ is sufficiently large. On the other
hand, in case 1 $K_{off}$ is chosen to be such that $K_{on}>
K_{off}$ for all $K_{on}$. Thus when $K_{on}$ increases, an
inclusion spends most of the time in the excited state if
$K_{on}>>K_{off}$. Thus at large $|G_{2}|$, the tendency is that
$D_{eff}$ first decreases as $K_{on}$ increases, then the
dependence becomes less significant at large $K_{on}$.

\subsection{Case 1}

In case 1, when $K_{off}\Delta t\gtrsim10^{-3}$ (i.e.,
$K_{off}\gtrsim 10^{2} s^{-1}$, characteristic time scale of a
cycle of a motor protein), no inclusion clusters are observed in
simulations for $K_{on}\lesssim 10^{3}  s^{-1}$. This indicates
that in a system with typical inclusion excitation and relaxation
rates, activities may completely suppress the formation of
inclusion clusters. This is because when $K_{off}$ is sufficiently
large, the lifetime of excited state inclusions is sufficiently
short such that the excited-state inclusions does not have time to
aggregate. For $K_{off}\Delta t=10^{-4}$ (i.e., $K_{off}\sim 10
s^{-1}$, this is more likely for the case when the relaxation of
an inclusion is induced by ligands), inclusion clusters are
observed for a wide range of $K_{on}$.

\begin{figure}[tbp]
\begin{center}
\rotatebox{0}{ \epsfxsize= 4.5 in \epsfbox{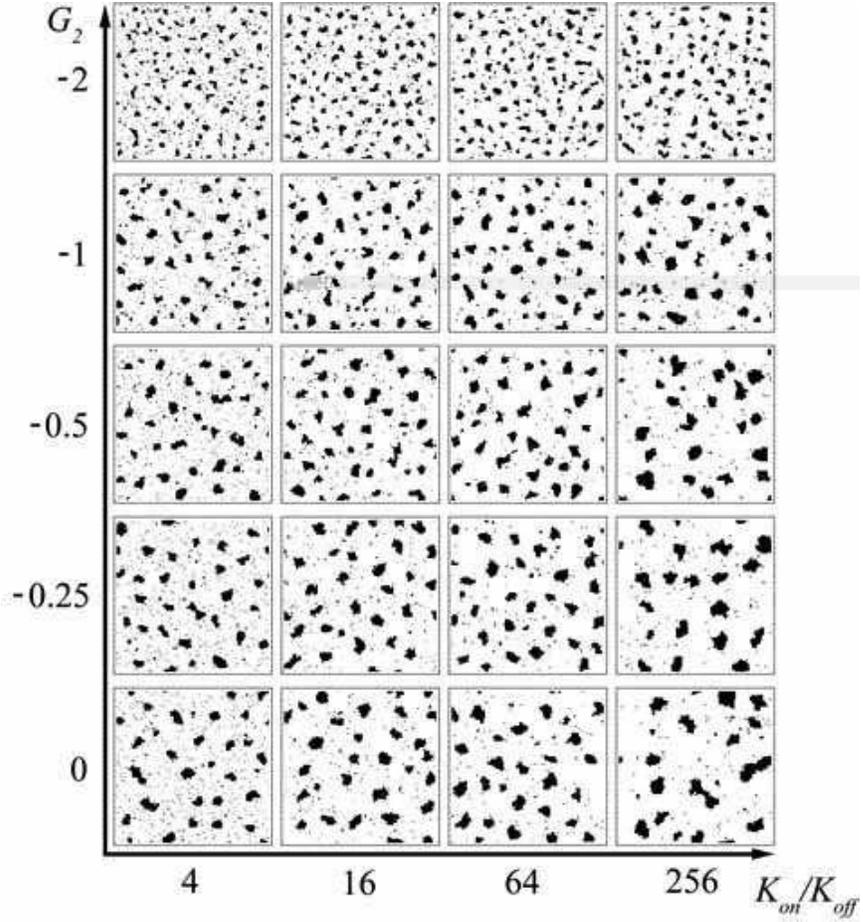} }
\caption{The distribution of inclusions in the membrane for case 1
with $\phi_{inc}=12.5\%$, $\widetilde{J}=1.5$, and $K_{off}\Delta
t=10^{-4}$. The typical length scale of inclusion clusters
increases as $K_{on}$ increases or $|G_{2}|$ decreases. Moreover,
the typical size of inclusion clusters increases faster when
$|G_{2}|$ is smaller.} \vskip 40 pt \label{fig:125_e}
\end{center}
\end{figure}

\begin{figure}[tbp]
\begin{center}
\rotatebox{0}{ \epsfxsize= 4.5 in \epsfbox{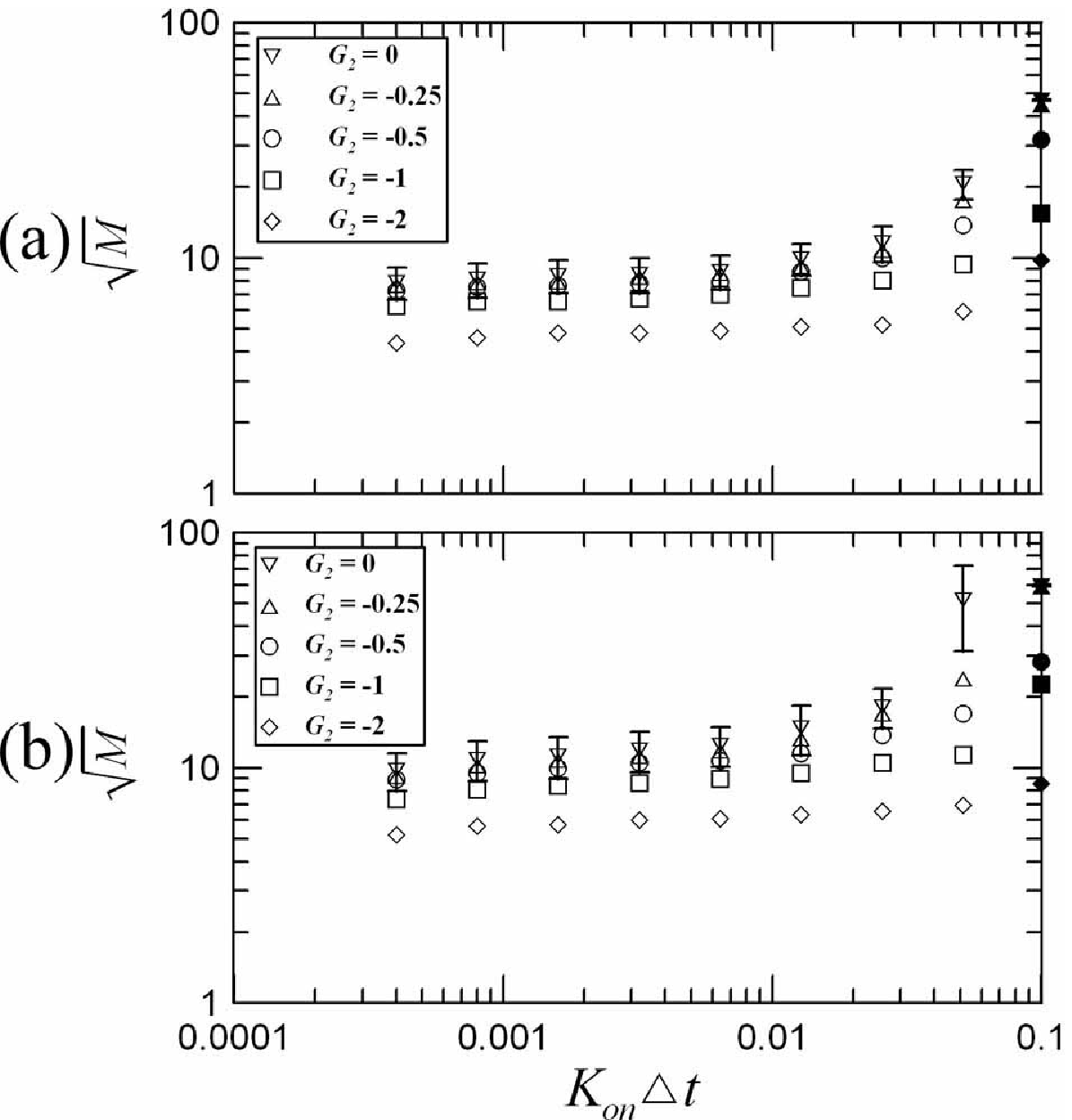} }
\caption{This relation between $\sqrt{M}$ and $K_{on}$ in case 1
for $K_{off}\Delta t=10^{-4}$, $G_{2}=0$ (triangle-down), $-0.25$
(triangle-up), $-0.5$ (circle), $-1$ (square), and $-2$ (diamond),
(a): $\phi_{inc}=12.5\%$, $\widetilde{J}=1.5$; (b):
$\phi_{inc}=25\%$, $\widetilde{J}=1$. The filled symbols denote
the situation when $K_{on}\Delta t=1$, the maximum excitation rate
can be simulated in the simulations, the error bar in this
situation is smaller than the symbol. When the inclusion-curvature
coupling is small, $\sqrt{M}$ shows weak dependence on
$K_{on}\Delta t$ for a wide range of $K_{on}\Delta t$, but
increases fast when $K_{on}\Delta t \gtrsim 0.01$. Furthermore,
the coupling between inclusion density and membrane curvature
provides the upper limit of the typical size of the inclusion
clusters, and suppresses the formation of large inclusion
clusters.} \vskip 40 pt \label{fig:125-e-size-1}
\end{center}
\end{figure}

Figure~\ref{fig:125_e} shows the snapshots of the distribution of
inclusions in the membrane for case 1 with $\phi_{inc}=12.5\%$,
$\widetilde{J}=1.5$, and $K_{off}\Delta t=10^{-4}$ on a
$128\times128$ square lattice.
The typical size of inclusion clusters increases as $K_{on}$
increases or $|G_{2}|$ decreases. Moreover, the dependence of the
typical size of inclusion clusters on the value of $K_{on}$ is more
evident for small $|G_{2}|$. Figure~\ref{fig:125-e-size-1} shows the
relation between $\sqrt{M}$ and $K_{on}\Delta t$ for $K_{off}\Delta
t=10^{-4}$ on a $128 \times 128$ square lattice. In
Figure~\ref{fig:125-e-size-1} (a) $\phi_{inc}=12.5\%$,
$\widetilde{J}=1.5$; in Figure~\ref{fig:125-e-size-1} (b)
$\phi_{inc}=25\%$, $\widetilde{J}=1$. The only exception is the
empty symbol with the greatest value of $\sqrt{M}$ in
Figure~\ref{fig:125-e-size-1} (b), which is taken from simulations
performed on a $256\times 256$ square lattice.
The filled symbols denote the situation when $K_{on}\Delta t=1$, the
maximum excitation rate can be simulated in our simulations.

Both Figure~\ref{fig:125-e-size-1} (a) and (b) indicate that when
$G_{2}$ is small, $\sqrt{M}$ shows weak dependence on
$K_{on}\Delta t$ for a wide range of $K_{on}\Delta t$, but
increases fast when $K_{on}\Delta t\gtrsim 0.01$. This can be
understood by comparing the diffusion length of a ground state
inclusion within its lifetime and the typical size of inclusion
clusters. Because the diffusion distance during the lifetime of a
ground state inclusion is $\approx\sqrt{4D_{eff}{K_{on}}^{-1}}$.
Therefore when $\sqrt{4D_{eff}{K_{on}}^{-1}}\lesssim\sqrt{M}$,
more and more ground state inclusions can not escape from an
inclusion cluster as $K_{on}$ increases. This ``positive
feedback'' of $K_{on}$ dependence of $\sqrt{M}$ is responsible for
the strong $K_{on}$ dependence of $\sqrt{M}$ at large $K_{on}$.
From the discussion of the previous subsection, $D_{eff}\approx0.8
\mu m^2/sec\approx0.8 a^2/\Delta t$, then the critical length
scale is $\sqrt{4D_{eff}{K_{on}}^{-1}}\approx {K_{on}}^{-1/2}$.
Therefore, $\sqrt{M}$ increases fast at $K_{on}\Delta t
\gtrsim0.01$ ($\sqrt{M}\approx 10$) in case 1.

Another important feature of Figure~\ref{fig:125-e-size-1} is that
for the filled symbols with $G_{2}=0$, in
Figure~\ref{fig:125-e-size-1} (a)
$\sqrt{N_{inc}}\approx\sqrt{M}\approx 45$; in
Figure~\ref{fig:125-e-size-1} (b) $\sqrt{N_{inc}}=64$ and
$\sqrt{M}\approx 60$, i.e., almost all inclusions aggregate in the
same inclusion cluster .

Moreover, when $G_{2}\neq 0$, $\sqrt{M}$ decreases as $|G_{2}|$
increases. This indicates that even when all inclusions are all in
the excited state, the coupling between inclusion density and
membrane curvature provides the upper limit of the typical size of
the inclusion clusters, and suppresses the formation of large
inclusion clusters. This is because the system has lower free
energy when the inclusions locate in the regions with greater
membrane curvature. Therefore, the membrane prefers to form many
mountain-like regions with inclusions aggregating on them, when
$G\neq0$.

\subsection{Case 2}
In case 2, inclusion clusters have been observed for
$K_{off}\Delta t=10^{-2} - 10^{-3}$, and a wide range of $K_{on}$.
All simulations are performed on a $128\times128$ square lattice
in case 2.

\begin{figure}[tbp]
\begin{center}
\rotatebox{0}{ \epsfxsize= 4.5 in \epsfbox{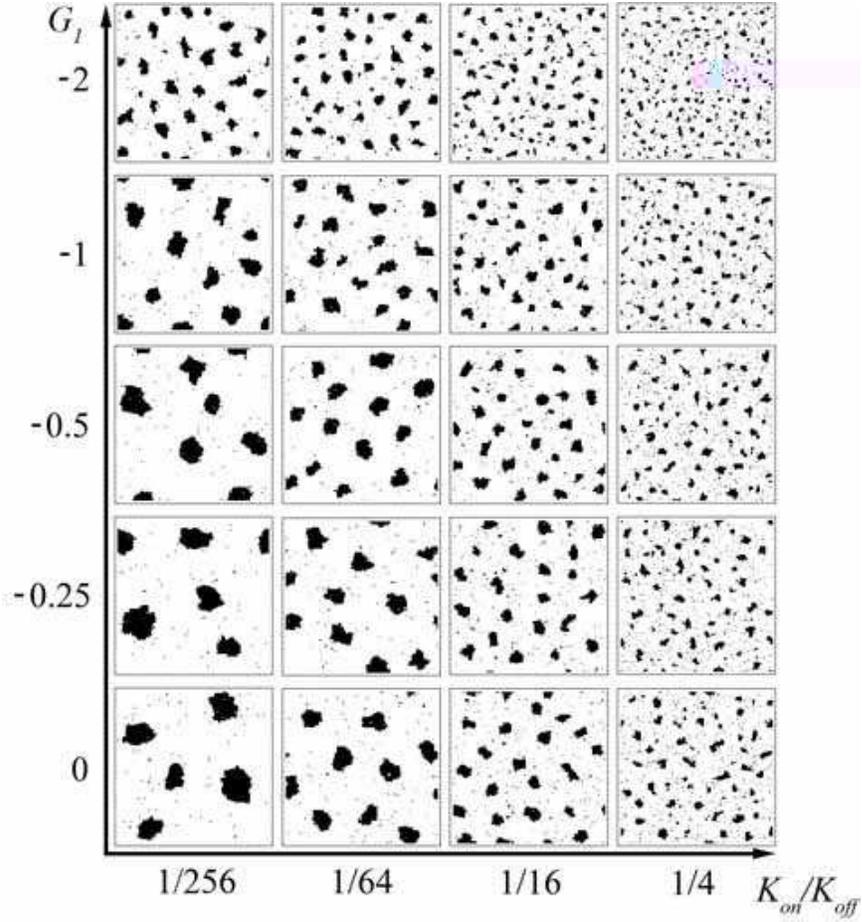} }
\caption{The distribution of inclusions in the membrane for case 2
with $\phi_{inc}=12.5\%$, $K_{off}\Delta t=10^{-3}$, and
$\widetilde{J}=1.5$. The typical length scale of inclusion cluster
increases as $K_{on}$ decreases or $|G_{1}|$ decreases. Moreover,
the typical size of inclusion clusters increases faster when
$|G_{1}|$ is smaller.} \vskip 40 pt \label{fig:125_g}
\end{center}
\end{figure}

\begin{figure}[tbp]
\begin{center}
\rotatebox{0}{ \epsfxsize= 4.0 in \epsfbox{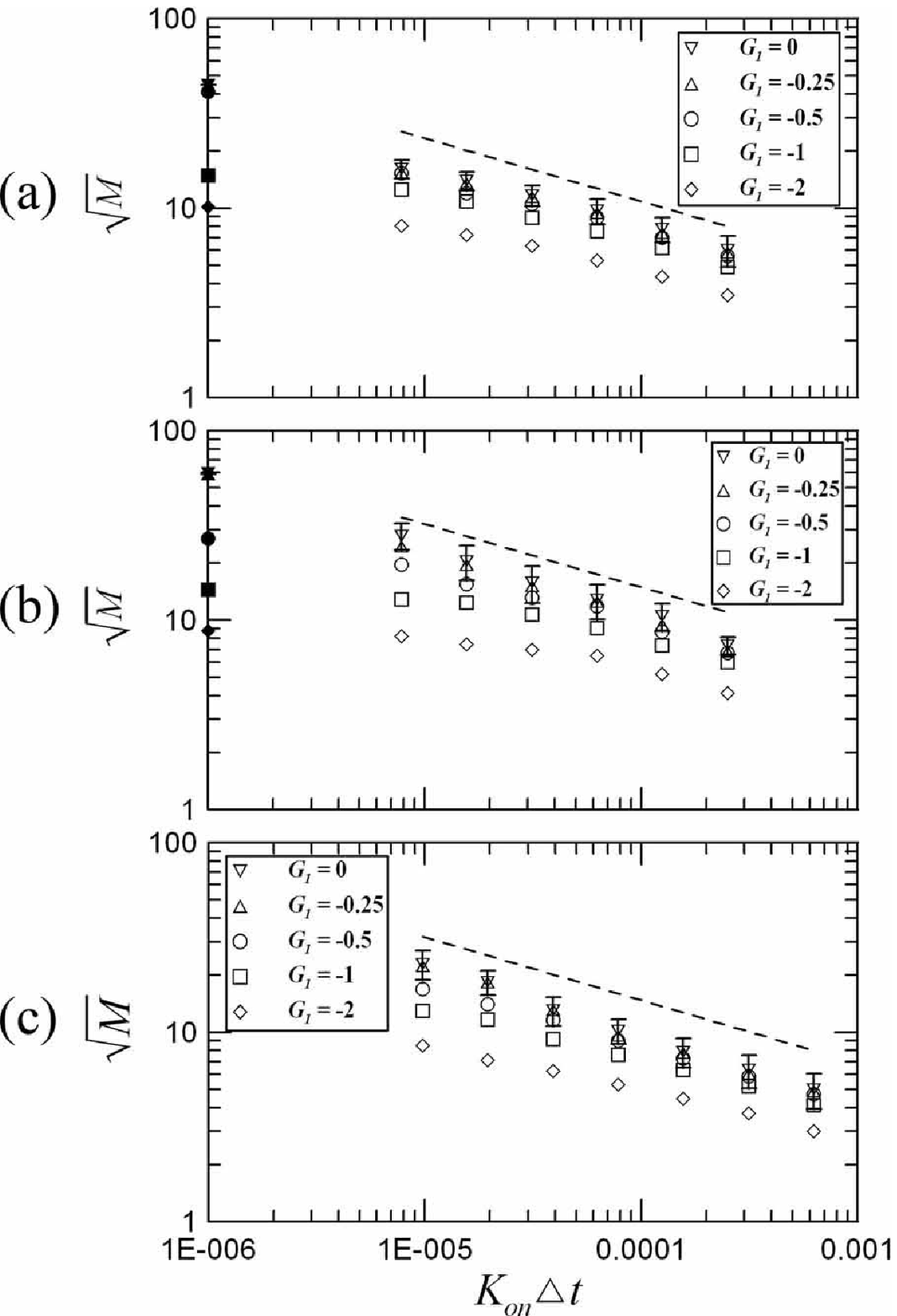} }
\caption{The relation between $\sqrt{M}$ and $K_{on}$ in case 1
for (a): $\phi_{inc}=12.5\%$, $\widetilde{J}=1.5$, and
$K_{off}\Delta t=10^{-3}$; (b): $\phi_{inc}=25\%$,
$\widetilde{J}=1$, and $K_{off}\Delta t=10^{-3}$; (c):
$\phi_{inc}=12.5\%$, $\widetilde{J}=1.5$, $K_{off}\Delta
t=10^{-2}$. The filled symbols denote the situation when all
inclusions are in the ground state, and $K_{on}=0$, and the error
bar in this situation is smaller than the symbol. $G_{1}=0$ case
agrees with $\sqrt{M}\sim {K_{on}}^{-\frac{1}{3}}$ relation pretty
well for a wide range of $K_{on}$. The slope of the dotted line is
$-\frac{1}{3}$.} \vskip 40 pt \label{fig:125-g-size-1}
\end{center}
\end{figure}

Figure~\ref{fig:125_g} shows the snapshots of the distribution of
inclusions in the membrane for case 2 with $\phi_{inc}=12.5\%$,
$\widetilde{J}=1.5$, and $K_{off}\Delta t=10^{-3}$.
The typical size of inclusion clusters decreases as $K_{on}$
increases or $|G_{1}|$ increases. Moreover, the dependence of the
typical size of inclusion clusters on the value of $K_{on}$ is more
evident for small $|G_{1}|$. Figure~\ref{fig:125-g-size-1} shows the
relation between $\sqrt{M}$ and $K_{on}\Delta t$. In
Figure~\ref{fig:125-g-size-1} (a) $\phi_{inc}=12.5\%$,
$\widetilde{J}=1.5$, and $K_{off}\Delta t=10^{-3}$,
in Figure~\ref{fig:125-g-size-1} (b) $\phi_{inc}=25\%$,
$\widetilde{J}=1$, and $K_{off}\Delta t=10^{-3}$, in
Figure~\ref{fig:125-g-size-1} (c) $\phi_{inc}=12.5\%$,
$\widetilde{J}=1.5$, and $K_{off}\Delta t=10^{-2}$. The filled
symbols denote the situation when all inclusions are in the ground
state, i.e., $K_{on}=0$.

When the density of the inclusions is not coupled to local
membrane curvature, it agrees with $\sqrt{M}\sim
{K_{on}}^{-\frac{1}{3}}$ relation pretty well for a wide range of
$K_{on}$. This is because when there is no budding in the
membrane, the growth of inclusion clusters in the absence of
inclusion activities and inclusion-curvature coupling corresponds
to a two-dimensional phase separation dynamics, i.e.,
$\sqrt{M}\sim t^{1/3}$.~\cite{ref:Seul_94, ref:Bray_94} This
growth eventually saturates due to the active transitions of the
inclusions. Since the lifetime of ground state inclusions is
${K_{on}}^{-1}$, thus the typical length scales of inclusion
clusters in the steady state should obey $\sqrt{M}\sim
{K_{on}}^{-1/3}$.~\cite{ref:Hsuan_04}

The same as case 1, for the filled symbol with $G_{1}=0$, almost
all inclusions aggregate in the same inclusion cluster (in
Figure~\ref{fig:125-g-size-1} (a)
$\sqrt{N_{inc}}\approx\sqrt{M}\approx 45$; in
Figure~\ref{fig:125-g-size-1} (b) $\sqrt{N_{inc}}=64$ and
$\sqrt{M}\approx 60$). Moreover, when $G_{1}\neq 0$, $\sqrt{M}$
decreases as $|G_{1}|$ increases. Therefore, the coupling between
inclusion density and membrane curvature provides the upper limit
of the typical size of the inclusion clusters, and suppresses the
formation of large inclusion clusters as in case 1.

Notice that Figure~\ref{fig:125-g-size-1} (a) and (b) have
different $\phi_{inc}$ and different $J_{mn}$, but at the same
$K_{on}$ the typical length scales of inclusion clusters are
almost in the same range. Furthermore, $K_{off}$ in (c) is tenfold
of $K_{off}$ in (a), but at the same $K_{on}$ the typical length
scales of inclusion clusters are also almost in the same range.
This means that $K_{on}$ is the key factor controls the size of
inclusion clusters.

In case 2 the lifetime of the state that prefers to aggregate is
changed when $K_{on}$ is changed, but in case 1 the lifetime of
the state that prefers to leave the inclusion clusters is changed
when $K_{on}$ is changed. Therefore, case 1 and case 2 are
different, and this is the reason why the $K_{on}$ dependence in
these two cases are so different.

\chapter{Summary \label{chapter 5}}

Our Monte Carlo simulations on a simple toy model have shown that
inclusion activities and inclusion-curvature coupling both could
contribute to the formation of finite-size inclusion clusters in
biological membranes. Since the inclusions have two internal
states, and the conformational changes of the inclusions are
induced by external stimuli. Thus, we treat a biomembrane as an
active, nonequilibrium system. Two special cases with different
interactions between the inclusions and lipids are studied in the
simulation. In case 1, excited state inclusions attract with each
other and induce local membrane curvature; ground state inclusions
do not attract or repel with other molecules, and the density of
ground state inclusions is not coupled to local membrane
curvature. In case 2, ground state inclusions tend to attract with
each other and induce local membrane curvature; excited state
inclusions do not attract or repel with other molecules, and the
density of excited state inclusions is not coupled to local
membrane curvature. Our main results are:

(1) The effective diffusion constant of the inclusions decreases
as the inclusion-curvature coupling and the lifetime of the state
with inclusion-curvature coupling increases.

(2) In case 1, we find that when $K_{off}$ is sufficiently large,
the excited-state inclusions does not have time to aggregate, and
there is no inclusion cluster in the system for a wide range of
$K_{on}$.

(3) In case 1, when inclusion-clusters do form, typical size of
clusters shows weak dependence on the excitation rate of the
inclusions for a wide range of $K_{on}$, but increases fast when
$K_{on}$ is greater than some critical value. Because the
diffusion distance of a ground state inclusion in the inclusion
cluster $\sim\sqrt{4D_{eff}{K_{on}}^{-1}}\sim {K_{on}}^{-1/2}$.
When $\sqrt{M}\gtrsim {K_{on}}^{-1/2}$ the inclusions become hard
to escape from an inclusion cluster.

(4) In case 2, we find that when the density of the inclusions is
not coupled to local membrane curvature, it agrees with
$\sqrt{M}\sim {K_{on}}^{-\frac{1}{3}}$ relation pretty well for a
wide range of $K_{on}$. This is because when there is no budding
in the membrane, the growth of inclusion clusters in the absence
of inclusion activities and inclusion-curvature coupling
corresponds to a two-dimensional phase separation dynamics, i.e.,
$\sqrt{M}\sim t^{1/3}$. This growth eventually saturates at time
scale $\sim {K_{on}}^{-1}$ due to the active transitions of the
inclusions.

(5) Moreover, in both case 1 and case 2, we find that the coupling
between inclusion density and membrane curvature provides the
upper limit of the typical size of the inclusion clusters, and
suppresses the formation of large inclusion clusters.

Our simple model has neglected many complications that occur in
real biological membrane. For example: (i) We ignored the effect
of cytoskeleton on the motion of the inclusions.~\cite{ref:Lin_04}
(ii) The effect of solution flow is not incorporated in our
model.~\cite{ref:Doi_book} (iii) Our model is not appropriate to
describe budding in biomembrane.~\cite{ref:HG_95} New models that
includes the aforementioned mechanisms will be our future research
direction.

\newpage
{\thispagestyle{empty} \topskip 3in
\begin{center}
{\Huge \textbf{APPENDICES}}
\end{center}}

\appendix
\chapter{Non-dimensionalization of the Hamiltonian}
The Hamiltonian of the membrane can be written as
\begin{align}\label{eq:Hami_appendix_01}
  H_{lattice}=&\sum_{(i,j)}\frac{a^2}{2}\left[\kappa{\left(\nabla^2_{\bot}h\right)^2}_{ij}+\gamma\left(\left|{\nabla}_{\bot}h\right|^2\right)_{ij}\right] \notag \\
  &- \sum_{(i,j)}\sum_{p=0}^{2}a^2\kappa C_{p}\phi_{p}(i,j)\left(\nabla^2_{\bot}h\right)_{ij} \notag \\
  &+
  \frac{1}{2}\sum_{\langle(i,j)(k,l)\rangle}\left(\sum_{m,n}J_{mn}\phi_{m}(i,j)\phi_{n}(k,l)\right).
\end{align}
First, we choose the unit of energy to be $k_{B}T$. Then we choose
the unite length of local vertical height of membrane to be
$h_{0}\sim \frac{a\sqrt{k_{B}T}}{\sqrt{\kappa}}$. Thus the first
term of $H_{lattice}$ at $h\sim h_{0}$,
${\nabla_{\bot}}^{2}\sim\frac{1}{a^2}$ is on the order of $k_{B}T$.
The discretized Laplacian and gradient of $h$ in two dimension are
\begin{align}\label{eq:Hami_appendix_05}
  \nabla^2_{\bot} h &= \frac{\partial^2 h}{\partial x^2}
+ \frac{\partial^2 h}{\partial y^2} \nonumber \\
&=  \frac{\partial}{\partial x}\left( \frac{h_{x + \frac{a}{2}, y} -
h_{x - \frac{a}{2}, y }}{a}\right) + \frac{\partial}{\partial
y}\left(\frac{h_{x, y + \frac{a}{2}} - h_{x,
y - \frac{a}{2}}}{a} \right) \nonumber \\
&= \frac{1}{a^2}( h_{x+a,y} + h_{x-a,y} + h_{x,y+a}+ h_{x,y-a} -
4h_{x,y})\nonumber \\
&= \frac{a}{a^2} \sqrt{\frac{k_{B}T}{\kappa }} \left(\widetilde{h}
_{i+1,j}+\widetilde{h}_{i-1,j}+\widetilde{h}_{i,j+1}+\widetilde{h}
_{i,j-1}-4\widetilde{h}_{i,j}\right)
\nonumber \\
&\equiv \frac{1}{a} \sqrt{\frac{k_{B}T}{\kappa}}
\widetilde{\nabla}^2_{\bot}\widetilde{h},
\end{align}
and
\begin{align}\label{eq:Hami_appendix_06}
{\nabla}_{\bot} h &= \frac{\partial h}{\partial x}\hat{i} +
\frac{\partial h}{\partial y}\hat{j} \nonumber \\
&= \frac{1}{2a} \left[ ( h_{x+a,y} - h_{x-a,y} )\hat{i} + (
h_{x,y+a} - h_{x,y-a})\hat{j} \right] \nonumber \\
&= \frac{a}{2a} \sqrt{\frac{k_{B}T}{\kappa}} \left[
\left(\widetilde{h}_{i+1,j} - \widetilde{h}_{i-1,j}\right)\hat{i} +
\left(\widetilde{h}_{i,j+1}-\widetilde{h}_{i,j-1}\right)\hat{j}
\right]
\nonumber \\
&\equiv \sqrt{\frac{k_{B}T}{\kappa}} \widetilde{{\nabla}}_{\bot}
\widetilde{h}.
\end{align}
Thus, total Hamiltonian becomes
\begin{align}\label{eq:Hami_appendix_07}
  k_{B}T\widetilde{H}_{lattice}=&\sum_{(i,j)}\frac{a^2}{2}\left[\kappa\frac{k_{B}T}{a^2\kappa}{\left(\widetilde{\nabla}^2_{\bot}\widetilde{h}\right)^2}_{ij}
  +\gamma\frac{k_{B}T}{\kappa}\left(\left|\widetilde{{\nabla}}_{\bot}\widetilde{h}\right|^2\right)_{ij}\right] \notag \\
  &- \sum_{(i,j)}\sum_{p=0}^{2}a^2\kappa C_{p}\phi_{p}(i,j)\frac{\sqrt{k_{B}T}}{a\sqrt{\kappa}}\left(\widetilde{\nabla}^2_{\bot}\widetilde{h}\right)_{ij} \notag \\
  &+
  \frac{1}{2}\sum_{\langle(i,j)(k,l)\rangle}\left(\sum_{m,n}k_{B}T\widetilde{J}_{mn}\phi_{m}(i,j)\phi_{n}(k,l)\right).
\end{align}
From the definition of dimensionless of inclusion-curvature
coupling $G_{i}\equiv h_{0}C_{i}$, we can obtain the
non-dimensionalized Hamiltonian of the system
\begin{align}\label{eq:Hami_appendix_08}
  \widetilde{H}_{lattice}=&\sum_{(i,j)}\frac{1}{2}\left[{\left(\widetilde{\nabla}^2_{\bot}\widetilde{h}\right)^2}_{ij}
  +\frac{\gamma a^2}{\kappa}\left(\left|\widetilde{{\nabla}}_{\bot}\widetilde{h}\right|^2\right)_{ij}\right] \notag \\
  &- \sum_{(i,j)}\sum_{p=0}^{2} G_{p}\phi_{p}(i,j)\left(\widetilde{\nabla}^2_{\bot}\widetilde{h}\right)_{ij} \notag \\
  &+
  \frac{1}{2}\sum_{\langle(i,j)(k,l)\rangle}\left(\sum_{m,n}\widetilde{J}_{mn}\phi_{m}(i,j)\phi_{n}(k,l)\right).
\end{align}

\newpage
{\thispagestyle{empty} \topskip 3in
\begin{center}
\Huge \bf BIBLIOGRAPHY
\end{center}}

\bigskip

\end{document}